\documentclass[aps,pra,floatfix,preprint,superscriptaddress,nofootinbib]{revtex4-1}
\usepackage[utf8]{inputenc}
\usepackage{amsmath}
\usepackage{amsfonts}
\usepackage{amssymb}
\usepackage{graphicx,color}
\usepackage{subcaption}
\usepackage{hyperref}
\newcommand{\s}{\sigma}
\newcommand{\g}{\gamma}
\newcommand{\oo}{\mathcal{O}}

\usepackage{adjustbox}
\usepackage{multirow}
\usepackage{hhline}
\begin{document}
\count\footins = 1000
\title{Operator thermalization vs eigenstate thermalization}
\author{Aleksandar Bukva}
\author{Philippe Sabella-Garnier}
\author{Koenraad Schalm}
\affiliation{Instituut-Lorentz and Delta-ITP, Universiteit Leiden, P.O. Box 9506, 2300 RA Leiden, The Netherlands}
%\captionsetup{justification=justified,singlelinecheck=false}

\begin{abstract}
We study the characteristics of thermalizing and non-thermalizing operators in integrable theories as we turn on a non-integrable deformation. Specifically, we show that $\sigma^z$, an operator that thermalizes in the integrable transverse field Ising model, has mean matrix elements that resemble ETH, but with fluctuations around the mean that are sharply suppressed. This suppression rapidly dwindles as the Ising model becomes non-integrable by the turning on of a longitudinal field. We also construct a non-thermalizing operator in the integrable regime, which slowly approaches the ETH form as the theory becomes non-integrable. At intermediate values of the non-integrable deformation, one distinguishes a perturbatively long relaxation time for this operator.
\end{abstract}

\maketitle
\tableofcontents

\section{Introduction}
The question of how closed, unitary quantum systems can (appear to) thermalize has long been at the heart of statistical mechanics. Recently, it has become more pressing because of its implications for real-life experiments \cite{doi:10.1146/annurev-conmatphys-031214-014548,Kaufman794,PhysRevX.7.041047,PhysRevLett.123.130601,labuhn2016tunable,turner2018weak} and, through holography for the black hole information paradox \cite{Dymarsky:2016ntg,Brehm:2018ipf,Romero-Bermudez:2018dim, Hikida:2018khg,Anous:2019yku,Nayak:2019khe,besken2019quantum}. The usual answer to the puzzle is the eigenstate thermalization hypothesis (``ETH'')\cite{PhysRevA.43.2046,1994PhRvE..50..888S,Yukalov:2012pi,DAlessio:2016rwt}. At its core, the statement is the following. Suppose that there is a regime where the matrix elements of an observable $\oo$ in the basis of energy eigenstates closely approximate the following form:
\begin{align}
\langle m|\oo|n\rangle \approx O(E) \delta_{mn} + e^{-S(E)/2} f(E,\omega) R_{mn}~, \nonumber \\ 
E=\frac{E_m+E_n}{2}~~,~~\omega=E_n-E_m~, \label{eq:eth-ansatz}
\end{align}
where $O(E)$ and $S(E)$ are the microcanonical expectation value of $\oo$ and entropy at energy $E$, $f(E,\omega)$ is a smooth function and $R_{mn}$ is a random matrix with zero mean and unit variance. Then, it can be shown that the long time average of the expectation value of $\oo$ in a superposition of energy eigenstates (such as a state produced by a quench in the Hamiltonian of the system) will approach its thermal expectation value, with the temperature set by the average energy of the initial state. The hypothesis is that in generic quantum theories with a large number of degrees of freedom, ``most'' observables have matrix elements approximately of this form and therefore the system will appear to thermalize. 

Nevertheless, while this is a sufficient condition for apparent thermalization, it is not \emph{necessary}. It is often argued that the validity of equation (\ref{eq:eth-ansatz}) for generic operators is a symptom of quantum chaos. However, it has been noticed in various contexts that even in \emph{free} systems certain (usually composite) operators can relax to a thermal state, at least at the level of linear response \cite{Grozdanov:2015nea,Amado:2017kgr, Parker:2018yvk,2019arXiv190508266M, Banerjee:2019iwd,Craps:2019pma, Sabella-Garnier:2019tsi}.  In \cite{Sabella-Garnier:2019tsi}, it was in fact shown that the thermal retarded Green's function of an operator in a free theory will generically decay exponentially in time unless the operator satisfies a particular no-go condition. That condition is:
\begin{equation}
|\langle m|\oo|n\rangle|^2=0 ~~ \text{unless} ~~E_n-E_m=F^{(\oo)}_{i}(P_n-P_m)~, \label{eq:no-go}
\end{equation}
where $E_{m,n}$ and $P_{m,n}$ are the energy and momentum of the states, $F^{(\oo)}_{i}(P)$ are (not-necessarily continuous) functions that depend on $\oo$, with $i$ an index that runs over a finite (system size-independent) range. By extension, such a statement should hold in any integrable theory, with a combination of the extensive set of conserved charges playing  the role of momentum in the above expression.

In a generic non-integrable theory, finding an operator satisfying this no-go condition is hard, if not impossible, since the momentum difference and energy difference between two states are a priori independent quantities. In integrable theories, the extensive number of conserved quantities makes finding operators that satisfy this condition easier. For example, in a free field theory the field itself obeys it, and one easily constructs others. However, it is also not hard to evade the no-go condition: any operator that involves two uncorrelated momentum modes will do so (for example, the square of a free field). As explained in \cite{Sabella-Garnier:2019tsi}, even free and integrable theories therefore have many operators that thermalize in linear response.

Of course, integrable field theories cannot be said to be chaotic for any reasonable definition of the word. In fact, their spectrum is highly regular. The fact that, at least at the level of linear response, many operators are sufficiently blind to this structure (as expressed by violating this no-go condition) and appear to thermalize is the idea that we have called \emph{operator thermalization}. This is in contrast with eigenstate thermalization in which it is the (lack of) structure of the spectrum itself that is responsible for thermalization.

In this note, we aim to determine the difference between these two ideas more concretely by studying thermalization in a one-dimensional quantum Ising chain. When only a transverse field is present, the model is integrable, whereas it is chaotic for a certain regime with both transverse and longitudinal fields. As an example of OTH, the local magnetization $\sigma^z$, which violates the no-go condition relaxes even in the integrable regime. We show that its matrix elements in the integrable theory take a form that is \emph{also} consistent with equation (\ref{eq:eth-ansatz}) provided we average over small energy windows. However, a detailed examination shows that the integrable structure of the spectrum is reflected in a non-Gaussian spectrum for $R_{mn}$. As we turn on the non-integrable deformation and transition to the chaotic regime, $R_{mn}$ becomes smoother and one observes a classic example of ETH.

We then compare this with the behaviour of a non-thermalizing operator $\Gamma$ in the integrable regime. By construction, this operator satisfies the no-go condition and does not relax. It therefore manifestly does not obey the ETH ansatz, even after averaging. Deforming the theory to the non-integrable regime slowly induces a violation of the no-go condition, and the operator then approaches a form compatible with equation ($\ref{eq:eth-ansatz}$). In contrast to $\sigma^z$, one clearly sees the onset of a long perturbative slowest-relaxation timescale (``mean free path'') in the system, that shortens as the degree of non-integrability is increased. 

We summarize our results in table \ref{table:summary}. For other work on ETH in the context of integrable theories, see \cite{Fioretto_2010,PhysRevLett.106.140405,PhysRevB.91.155123,PhysRevB.97.035129,Byju:2018eyb,Banerjee:2019ilw}. 

\begin{table}
\begin{tabular}{|c||c|c|c|c|} \hline 
Theory & Operator & \multicolumn{1}{|p{2.5cm}|}{Satisifies no-go condition?} & Relaxes? & Obeys ETH ansatz? \\ \hline \hline
\multirow{2}{*}{\rotatebox[origin=bl]{90}{Integrable~~~~~~~}} & $\sigma^z$ & No & Yes& \multicolumn{1}{|p{3cm}|}{Yes, but with $R_{mn}$ more sharply peaked than a Gaussian} \\
& $\Gamma$ & Yes & No & No \\ \hline
\multirow{3}{*}{\rotatebox{90}{Non-integrable~~~~~}} & $\sigma^z$ & No & Yes& Yes \\ & & & & \\
& $\Gamma$ & No & \multicolumn{1}{|p{3cm}|}{Yes, but with a long relaxation time} & \multicolumn{1}{|p{3cm}|}{Yes, with $f(E,\omega)$ flat as a function of $\omega$} \\ \hline
\end{tabular}
\caption{Summary of results}
\label{table:summary}
\end{table}

\section{Model details}
The one-dimensional Ising model with transverse and longitudinal fields has the following Hamiltonian:
\begin{equation}
H=-J\sum_{n=1}^N\left(\sigma^{z}_i \sigma^{z}_{i+1}+h\sigma_i^x +g\sigma_i^z\right)~,
\end{equation}
where $\sigma_i^a$ are the usual Pauli matrices, obeying
\begin{equation}
[\s_i^a,\s_j^b]=2i\epsilon^{ab}_{~~c} \sigma_i^c \delta^{ij}~.
\end{equation}
We impose periodic boundary conditions, $\sigma^{a}_{i+N}\equiv \sigma^{a}_i$. When $g=0$, the transverse field Ising model is integrable and can be mapped to a model of free spinless fermions through a series of textbook transformations. First, a Jordan-Wigner transformation and Fourier transform will make the Hamiltonian quadratic:
\begin{align}
&\s_i^x=1-2c_i^\dagger c_i ~~,~~ \s_i^z=-\prod_{j<i}(1-2c_j^\dagger c_j)(c_i+c_i^\dagger)~, \nonumber\\
&c_j=\frac{1}{\sqrt{N}} \sum_{k \in \mathcal{K}} c_k e^{ikr_j} ~, \nonumber\\
&\{c_k,c^\dagger_{k'}\}=\delta_{kk'}~.
\end{align}
We think of the system as being on a lattice with lattice spacing $a$ and total size $L=Na$, so that $r_j=j a \in [a,N a]$. The periodic boundary conditions on $\s_i^a$ impose either periodic or anti-periodic boundary conditions on the fermionic operators (depending on the total number of fermions), leading to
\begin{equation}
\mathcal{K}= \left\{\frac{2\pi}{L} n \left| n\in \mathbb{Z} ~~\text{or}~~ \left(\mathbb{Z}+\frac{1}{2}\right) \right. \right\}~.
\end{equation}
Of course, the momenta must lie in the first Brillouin zone, leading to $-\frac{\pi}{a}<k\leq\frac{\pi}{a}$, so that $n \in \left ( -\frac{N}{2}, \frac{N}{2} \right ] $. In practice, we will work with $a=1$, so that $N$ measures system size. 
This transformation is followed by a Bogoliubov transformation:
\begin{align}
&c_k=u_k \g_k + iv_k \g^\dagger_{-k}~, \nonumber \\
&u_k=\cos(\theta_k/2)~,~v_k=\sin(\theta_k/2)~,~\tan\theta_k\equiv \frac{\sin(ka)}{h-\cos(ka)}~.
\end{align}
In terms of these fermions, the Hamiltonian is diagonal:
\begin{align}
H|_{g=0}=\sum_{k\in \mathcal{K}} \epsilon_k\left(\g^\dagger_k \g_k - \frac{1}{2}\right) ~,~~ \epsilon_k\equiv 2J\sqrt{1+h^2-2h\cos(ka)}~. 
\end{align}
The momentum operator is then
\begin{equation}
P=\sum_{k \in \mathcal{K}} k \g^\dagger_k \g_k~.
\end{equation}
Numerically, we work in this fermion basis, labelling states by occupation number of each of the momenta in $\mathcal{K}$ with the appropriate boundary conditions. To construct operators in the non-integrable regime, we first construct them in the basis of eigenstates of the integrable Hamiltonian. We then diagonalize the non-integrable Hamiltonian and numerically find the transformation between the eigenvectors. For the integrable model, we work in a basis of joint eigenvectors of the occupation number of each of the momentum modes. Away from integrability, the Hamiltonian is still translationally-invariant. We therefore work in a basis of joint eigenvectors of $H$ and the translation operator. 
Throughout this paper, we set the value of the transverse field to $h=-1.05$, following \cite{PhysRevLett.106.050405} which studied thermalization in the mixed field Ising chain. We will mostly focus on three values of the parallel field: $g=0$ (integrable), $g=0.1$ (which we label simply ``non-integrable'') and $g=0.5$, which (following \cite{PhysRevLett.106.050405}) we label ``far from integrable''. 

\begin{figure}
\begin{tabular}[c]{ccc}
\begin{subfigure}[c]{0.30\textwidth}
\includegraphics[width=\textwidth]{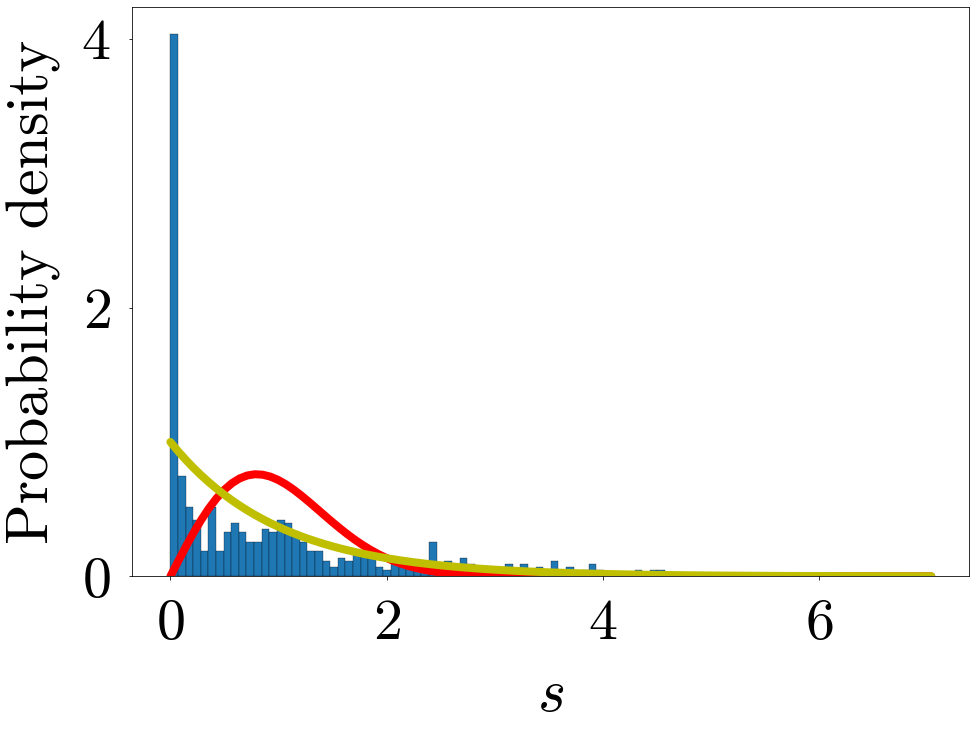}
\caption{Integrable ($g=0$)}
\end{subfigure} &
\begin{subfigure}[c]{0.30\textwidth}
\includegraphics[width=\textwidth]{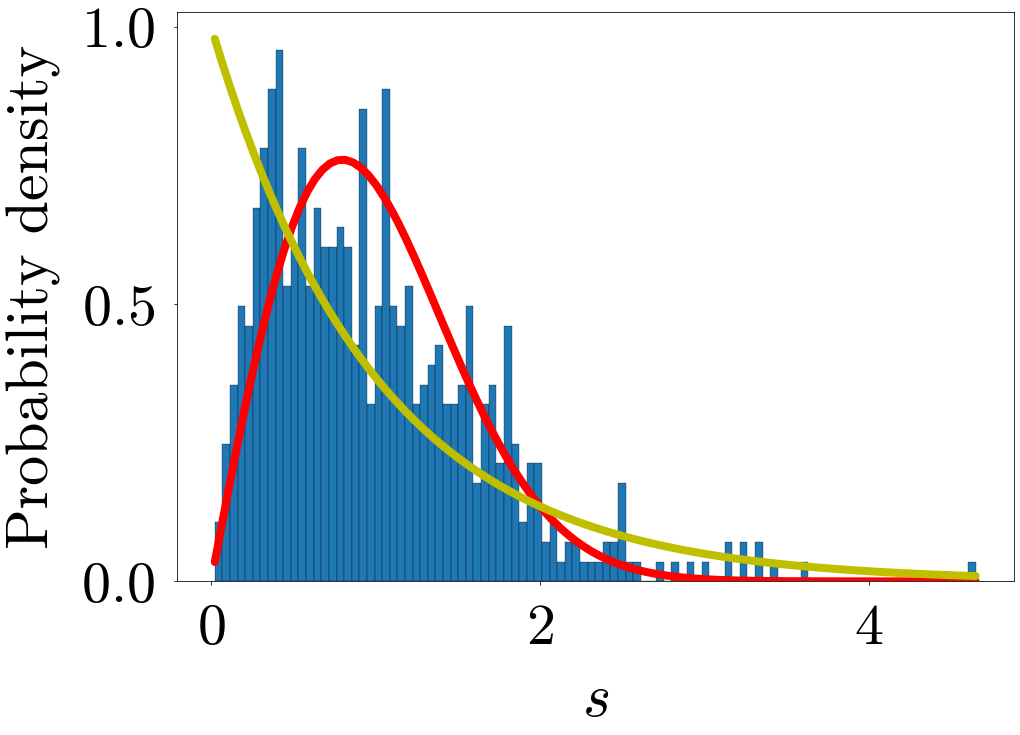}
\caption{Non-integrable ($g=0.1$)}
\end{subfigure}
\begin{subfigure}[c]{0.30\textwidth}
\includegraphics[width=\textwidth]{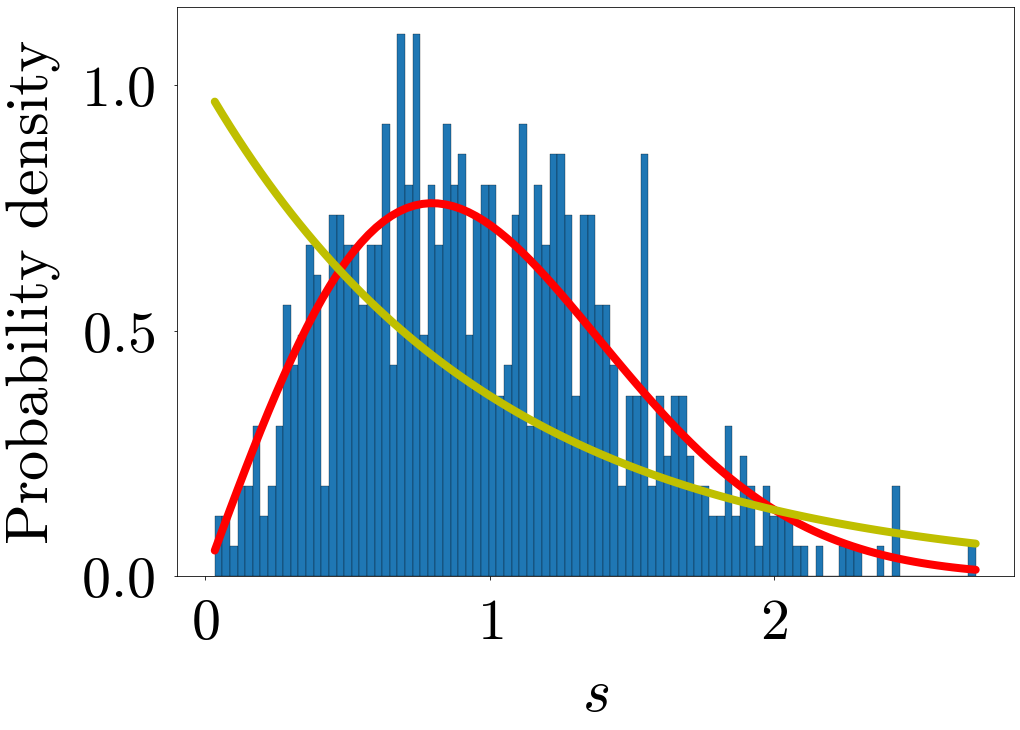}
\caption{Far from integrable ($g=0.5$)}
\end{subfigure}
\end{tabular}
\caption{(Colour online) Distribution of unfolded level spacings in units of average level spacing $s$ for states with one eigenvalue of the translation operator ($e^{\frac{(N-2)\pi i}{N}}$). The red line corresponds to the Wigner sumrise for the Gaussian Orthogonal Ensemble. The green line corresponds to a Poisson distribution. Level repulsion is very clearly visible for the far from integrable case. Note that the intermediate $g=0.1$ case  can be seen to approach  the Wigner sumrise overall but still shows an excess of approximate degeneracies. $N=13$.} 
\label{fig:level-stats}
\end{figure}

In figure \ref{fig:level-stats}, we show the level statistics for these three values of the transverse field in one particular sector (i.e. for states with one particular eigenvalue of the translation operator), confirming that the far-from-integrable case follows a Wigner distribution while the integrable case is Poisson-distributed.

\section{Operators: thermalizing and non-thermalizing}
\begin{figure}
\begin{tabular}[c]{cc}
\begin{subfigure}[c]{0.5\textwidth}
\includegraphics[width=\textwidth]{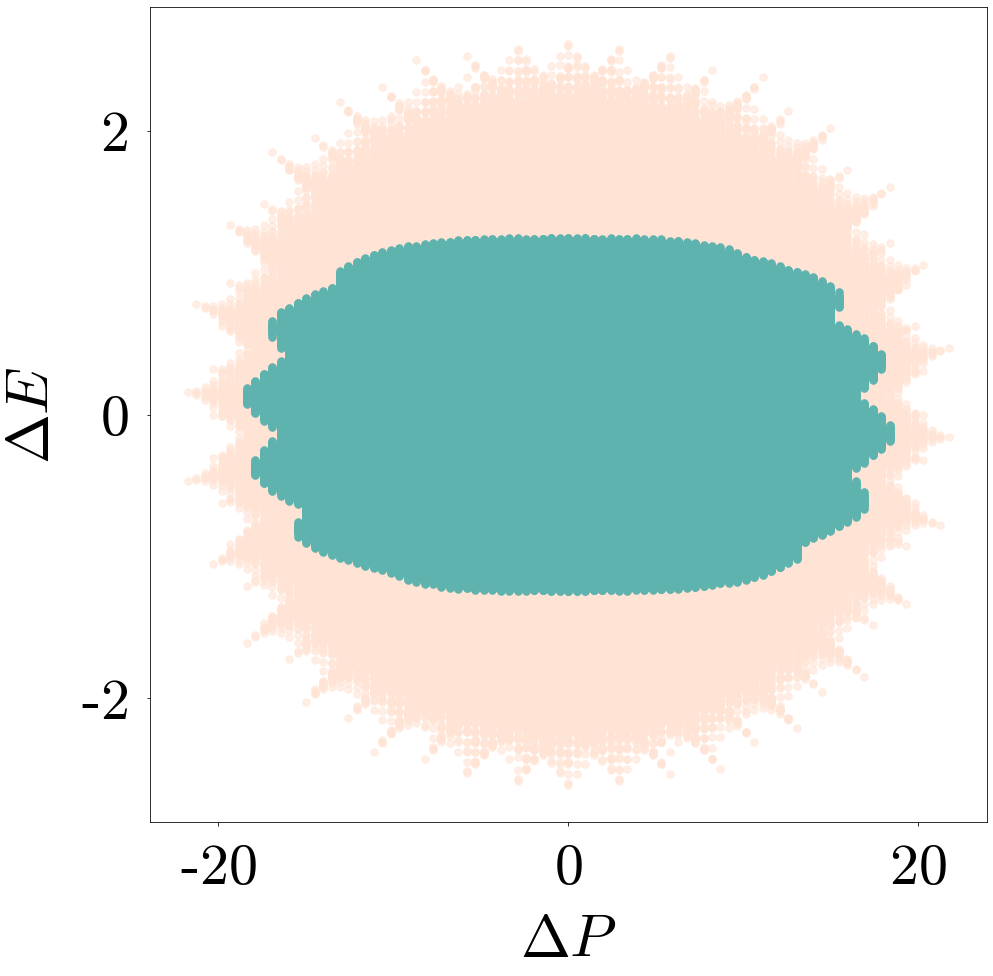}
\caption{$\sigma^z$}
\label{fig:no-go-sigma}
\end{subfigure} &
\begin{subfigure}[c]{0.5\textwidth}
\includegraphics[width=\textwidth]{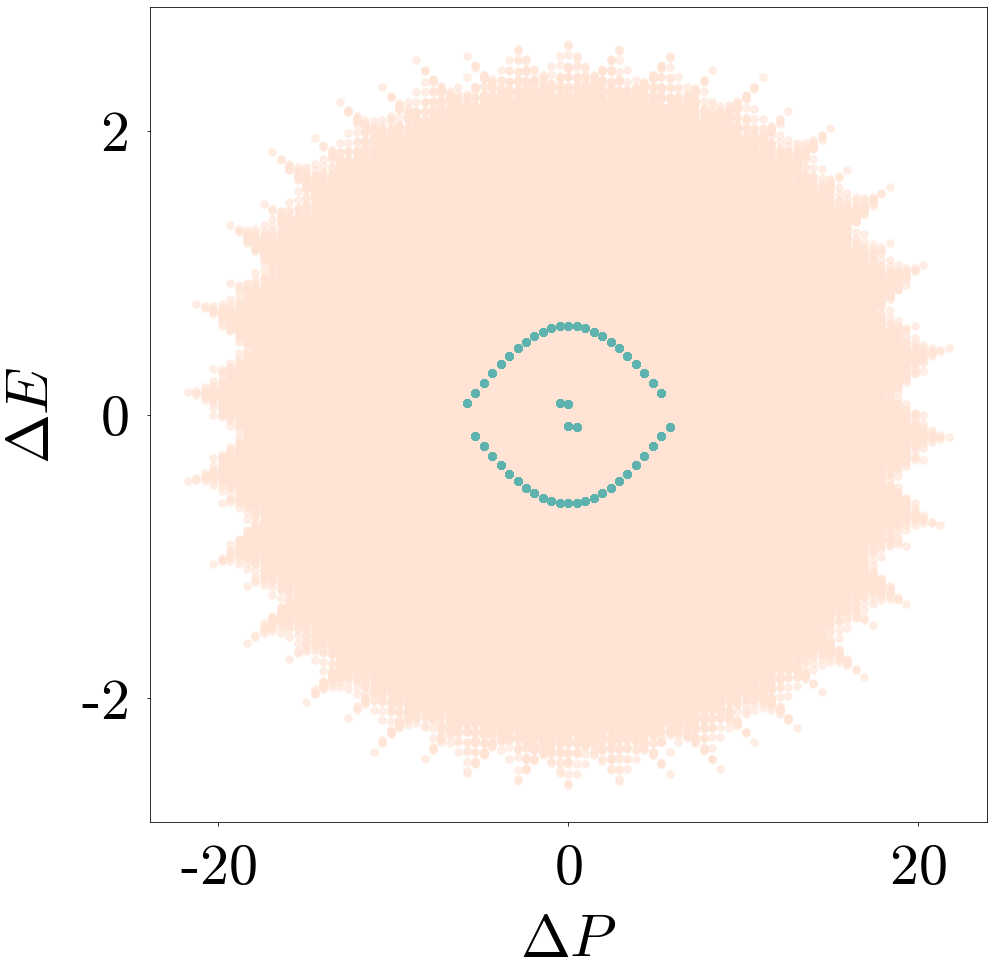}
\caption{$\Gamma$}
\label{fig:no-go-gamma}
\end{subfigure}
\end{tabular}
\caption{(Colour online) The no-go condition (\ref{eq:no-go}) for a thermalizing (\ref{fig:no-go-sigma}) and non-thermalizing (\ref{fig:no-go-sigma}) operator. Peach points correspond to zero matrix elements while teal points correspond to non-zero matrix elements. We can clearly see that for the thermalizing operator, the matrix elements are generically non-zero whereas for the non-thermalizing operator they are only non-zero when $\Delta E$ is given by a finite number of functions of $\Delta P$. $N=13$. Note that the fact that there is a non-zero matrix element with a particular $\Delta E$ and $\Delta P$ does not exclude that some other matrix element with those same values is zero.}
\end{figure}

As discussed in the introduction, we will be considering two different operators in both the integrable and non-integrable regimes. The first operator is $\sigma^z_i$. This operator clearly violates the no-go condition (\ref{eq:no-go}) in the integrable regime: its matrix elements are non-zero for a two-dimensional subregion of the $(\Delta E,\Delta P)$ plane, as opposed to a discrete set of lines. This is seen explicitly in figure \ref{fig:no-go-sigma} This confirms the analysis made in \cite{Sabella-Garnier:2019tsi} by analytical methods at the critical point ($h=1$).

By contrast, we can use the free fermion basis to construct an operator that obeys the no-go condition (\ref{eq:no-go}). Take 
\begin{equation}
\Gamma=\frac{2\pi}{L}\sum_{k \in \mathcal{K}} \left(\g_k \g_{k+\delta} +  \g^\dagger_{k+\delta} \g^\dagger_k\right)~,
%i\sum_{k \in \mathcal{K}}\left(e^{ikx} \g_k + e^{-ikx} \g_k^\dagger\right)%\left(e^{i(k+\delta)x}\g_{k+\delta}+e^{-i(k+\delta)x}\g^\dagger_{k+\delta}\right)~,
\end{equation}
where $\delta$ is an arbitrary (fixed) shift in momentum space. We will take it to be as small as possible, that is to say $\delta=\frac{2\pi}{N}$. This is the simplest operator that satisfies the no-go condition without being a conserved current. It creates pairs of particles with correlated momenta. It is easy to see that, in the integrable theory, such an operator has non-zero matrix elements only between states where
\begin{align}
&\Delta P= \pm (2k+\delta) \\
&\Delta E=\pm(\epsilon_k+\epsilon_{k+\delta})=\pm\left(\epsilon_{\frac{\Delta P-\delta}{2}}+\epsilon_{\frac{\Delta P+\delta}{2}}\right)~.
\end{align}
This is confirmed by figure \ref{fig:no-go-gamma}.

In the integrable regime, $\Gamma(x,t)$ is easily obtained by Fourier transforming:
\begin{equation}
\Gamma(x,t)=\frac{2\pi}{L}\sum_{k \in \mathcal{K}} \left(e^{i(2k+\delta)x} e^{-i(\epsilon_k+\epsilon_{k+\delta})t}\g_k \g_{k+\delta} + e^{-i(2k+\delta)x} e^{i(\epsilon_k+\epsilon_{k+\delta})t} \g^\dagger_{k+\delta} \g^\dagger_k\right)~,
\end{equation}

In the non-integrable regime, we can construct $\Gamma(x,t)$ by evolving $\Gamma$ in time explicitly with the non-integrable Hamiltonian and translation operator. 

\begin{figure}
\begin{tabular}[c]{cc}
\begin{subfigure}[c]{0.5\textwidth}
\includegraphics[width=\textwidth]{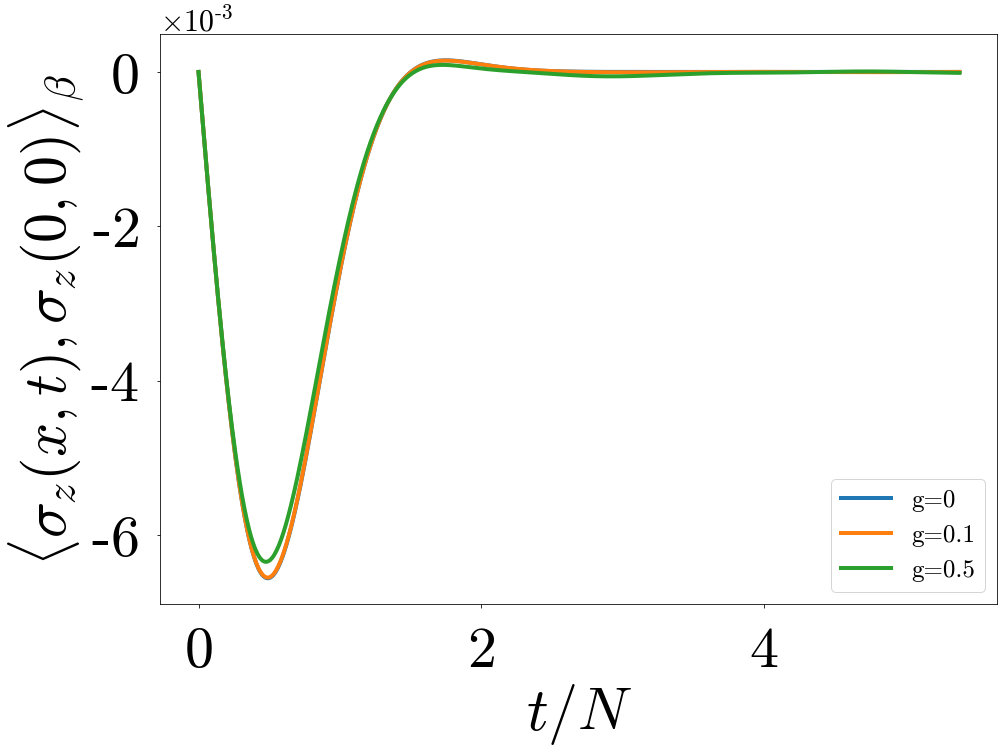}
\caption{$\sigma^z$}
\label{fig:gr-sigma}
\end{subfigure} &
\begin{subfigure}[c]{0.5\textwidth}
\includegraphics[width=\textwidth]{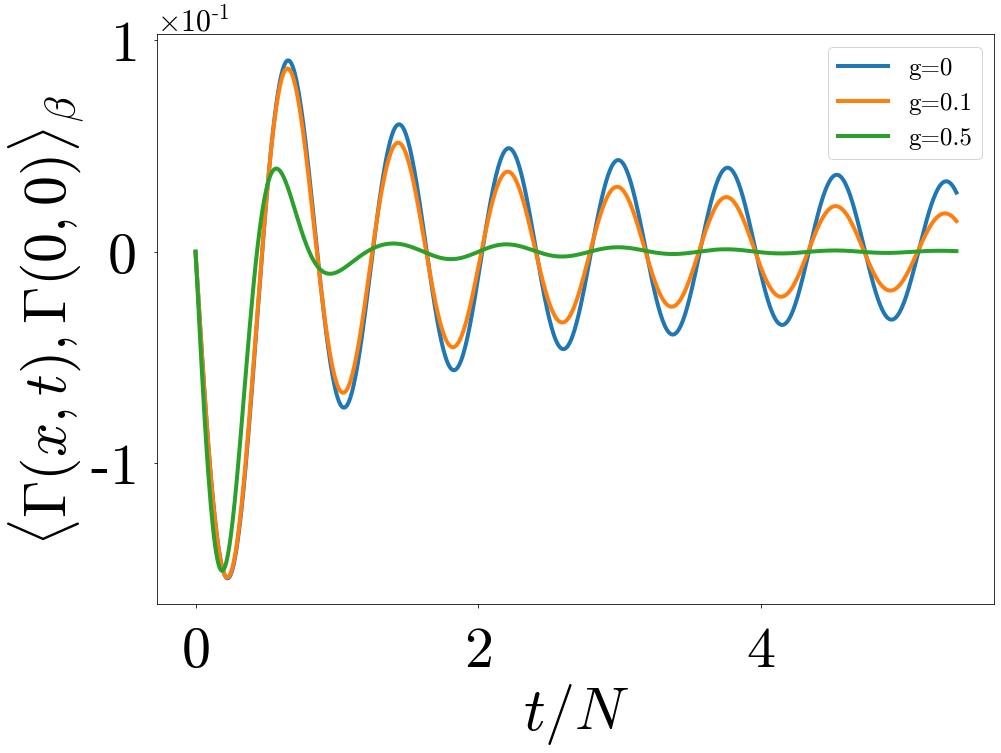}
\caption{$\Gamma$}
\label{fig:gr-gamma}
\end{subfigure}
\end{tabular}
\caption{Finite-temperature retarded Green's function for $\sigma^z$ and $\Gamma$ for various integrability-breaking parameters $g$, with $h=-1.05$ We can see that $\sigma^z$ always relaxes whereas in the integrable regime $\Gamma$ does not, and becomes more damped as $g$ is increased. $N=13$}
\end{figure}

\begin{figure}
\begin{tabular}[c]{cc}
\begin{subfigure}[c]{0.5\textwidth}
\includegraphics[width=\textwidth]{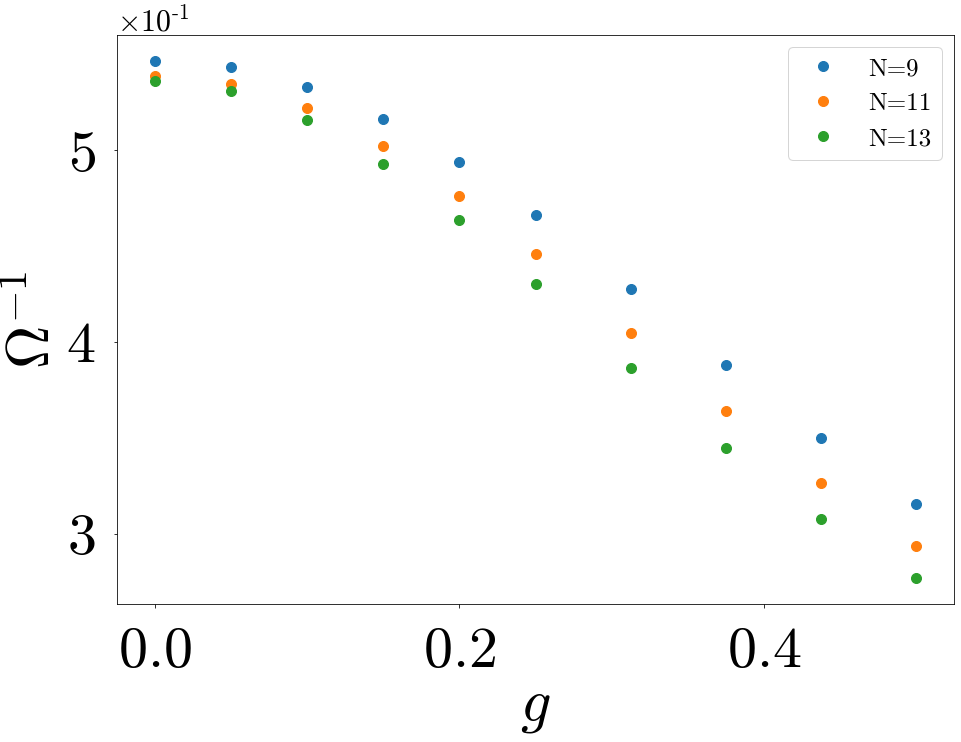}
\caption{Lifetime}
\end{subfigure} &
\begin{subfigure}[c]{0.5\textwidth}
\includegraphics[width=\textwidth]{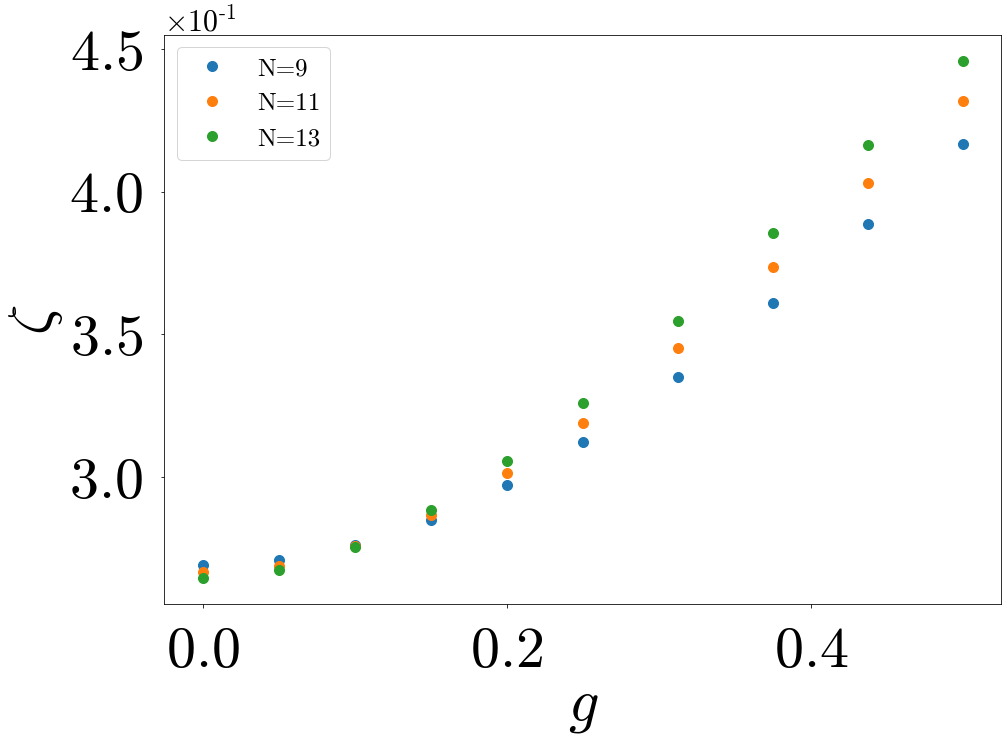}
\caption{Damping ratio}
\end{subfigure}
\end{tabular}
\caption{(Colour online) Lifetime (in units of system size) and associated damping ratio of an excitation of the thermal state by $\Gamma$ as a function of longitudinal field $g$ for different system sizes. We can see that for small $g$ the lifetime scales with system size. As $g$ is increased, the lifetime drops (and does so more steeply as we approach a continuum limit. We single out three values of $g$ that are of interest: at $g=0$ the theory is exactly integrable. $g=0.1$ displays a measurable break from integrability. Finally, $g=0.5$ is chaotic}
\label{fig:life-damp}
\end{figure}

In figure \ref{fig:gr-sigma}, we examine the finite-temperature retarded two-point function of $\sigma^z$, $-i\Theta(t)\langle[\sigma_i^z (t),\sigma_i^z(0)]\rangle_\beta$, as a function of time in both the integrable and non-integrable regimes. We can clearly see that it relaxes in both cases. To confirm that $\Gamma$ does not relax in the integrable theory, but does as we move away from integrability, we study its retarded Green's function as a function of the parameter $g$ while holding $h$ fixed. This is shown in figure \ref{fig:gr-gamma}. There are two characteristic timescales present in this response. We see that at $g=0$, the two-point function for $\Gamma$ does not relax, but as we increase $g$ it does. At $g=0.5$, it relaxes in a comparable manner to $\sigma^z$. We can Fourier transform $G_R(t)$ to better study the two timescales involved: the resulting frequency distribution can be fit to Lorentzian distributions, consistent with a signal of the form $e^{-\Omega t} \sin(\omega_0 t)$. The position of the peaks of the Lorentzian gives $\omega_0$ and their width gives $\Omega$. The lifetime of the excitation, $\Omega^{-1}$, and the damping ratio $\zeta=\sqrt{\frac{\Omega^2}{\omega_0^2+\Omega^2}}$ are shown in figure \ref{fig:life-damp} as a function of the magnitude of the longitudinal field. 

\section{OTH vs ETH}
Both OTH and ETH are fundamentally formulated in terms of the matrix elements of an operator in a basis of energy eigenstates, $\langle m|\oo|n\rangle$. OTH specifically is a corollary to the no-go condition expressed for $\langle m|\oo|n\rangle$. To study the relation between---and transition from--- OTH to ETH, we study the matrix elements of both $\sigma^z$ and $\Gamma$ as we turn on the non-integrable longitudinal field. Note that in the integrable theory, the energy eigenvalues are degenerate, whereas they are not so in the non-integrable theory (up to momentum). For a proper comparison, we will therefore at various stages take an average in both cases over a small energy window ($\delta E=0.01$) while holding $\omega$ fixed or vice-versa ($\delta \omega=0.8$).
\subsection{Thermalizing operator, $E$ dependence}
\begin{figure}
\begin{tabular}[c]{cc}
\begin{subfigure}{0.5\textwidth}
\includegraphics[width=\textwidth]{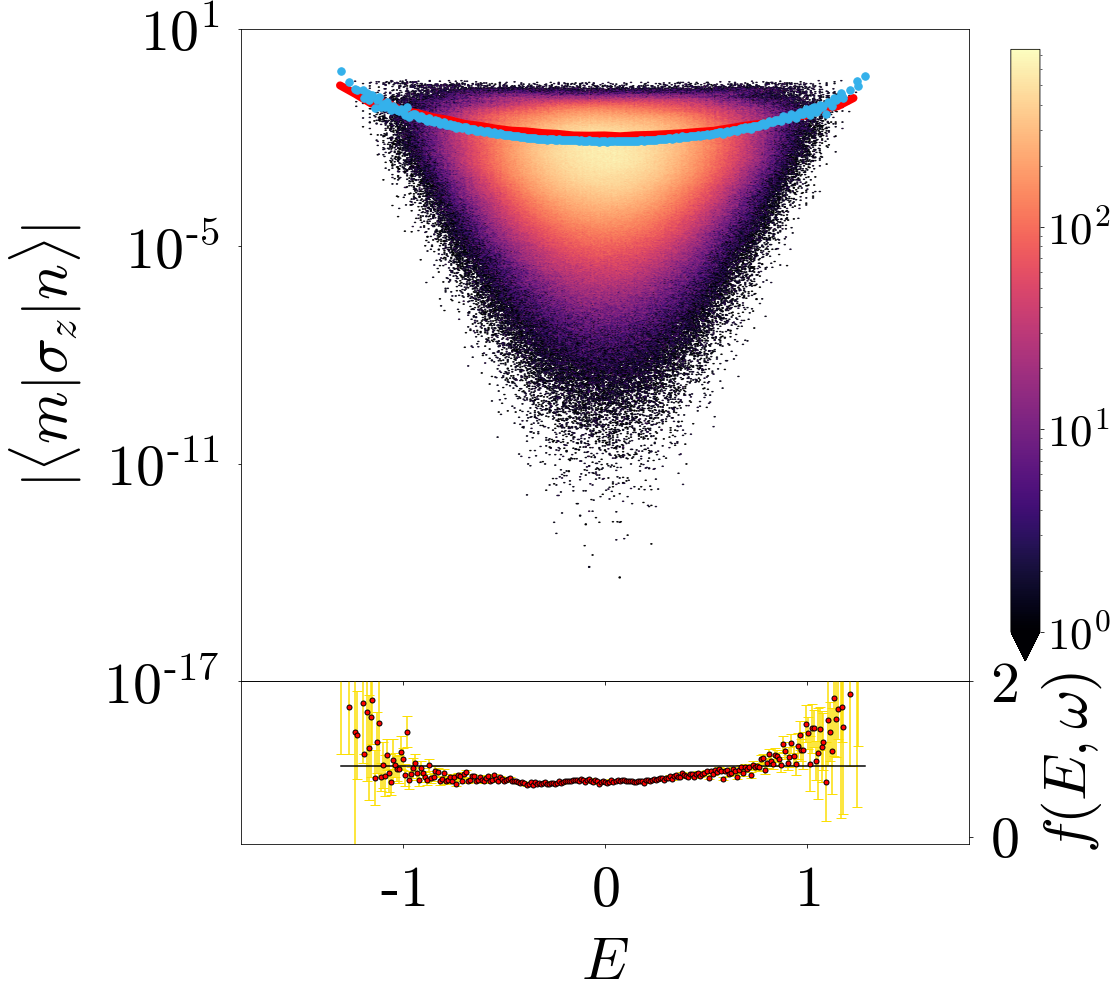}
\caption{Integrable}
\end{subfigure} &
\begin{subfigure}[r]{0.5\textwidth}
\includegraphics[width=\textwidth]{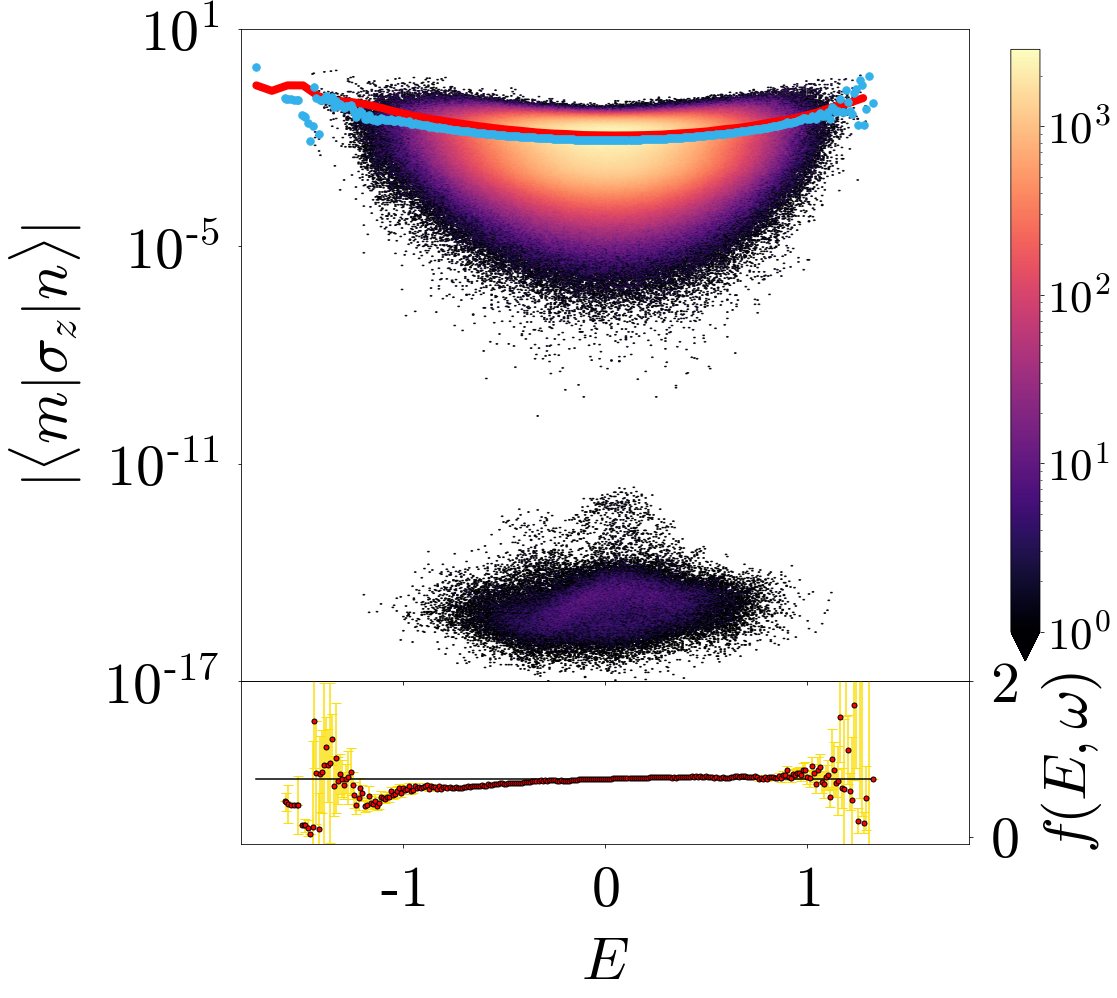}
\caption{Far from integrable}
\end{subfigure}
\end{tabular}
\caption{(Colour online) Dependence of the absolute value of matrix elements of $\sigma^z$ on the average energy of the states at fixed $\omega$. The blue points correspond to a running average over a small energy window and the red line is $e^{-S(E)/2}$. The bottom of the figure shows the running average divided by $e^{-S(E)/2}$, which gives the $E$ dependence of $|f(E,\omega)|$. In both cases, the result is consistent with that function not depending on $E$. The error bars correspond to a 95\% confidence interval if the underlying distribution is normal. \textcolor{red}{de}}
\label{fig:edep-sigma}
\end{figure}

In figure \ref{fig:edep-sigma}, we show the dependence of matrix elements of the thermalizing operator on the average energy of the states, $E=\frac{E_m+E_n}{2}$. In the integrable regime, half of the matrix elements $\langle m|\sigma^z|n\rangle$ are exactly zero because of parity symmetry (i.e. because the Hamiltonian is invariant under $\sigma_i^z \rightarrow -\sigma_i^z$). We exclude these points from our analysis. The behaviour of the remaining matrix elements in the integrable theory is strikingly similar to those in the non-integrable one. Taking a running average over a small (but finite) energy window allows us to extract a smooth function. In the non-integrable case, that should correspond to $e^{-S(E)/2} f(E,\omega)$. The same also happens---perhaps surprisingly--- in the integrable case: the average over an energy window also scales predominantly as $e^{-S(E)/2}$. This need not have been, but shows explicitly the similarity between OTH and ETH at the level of averages. It explains in particular why many studies in 2D CFTs, which have an extensive number of conserved quantities, nevertheless find ETH-like behaviour, even though it is usually a case of OTH (see, eg \cite{Brehm:2018ipf,Romero-Bermudez:2018dim,Hikida:2018khg,besken2019quantum}). We can see that most of the dependence on $E$ comes from this exponential factor of entropy, as expected. In the insert, we extract the function $f(E,\omega)$. 

\subsection{Non-thermalizing operator, $E$ dependence}
\begin{figure}
\begin{tabular}[c]{ccc}
\begin{subfigure}{0.33\textwidth}
\includegraphics[width=\textwidth]{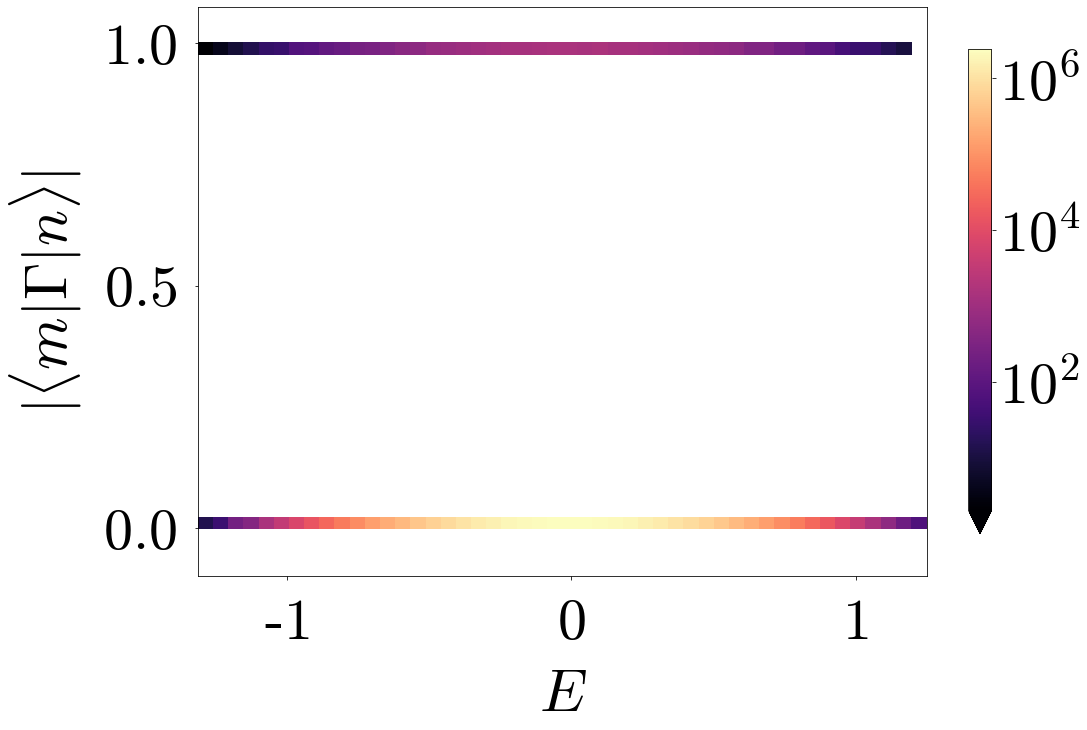}
\caption{Integrable}
\end{subfigure} &
\begin{subfigure}[r]{0.33\textwidth}
\includegraphics[width=\textwidth]{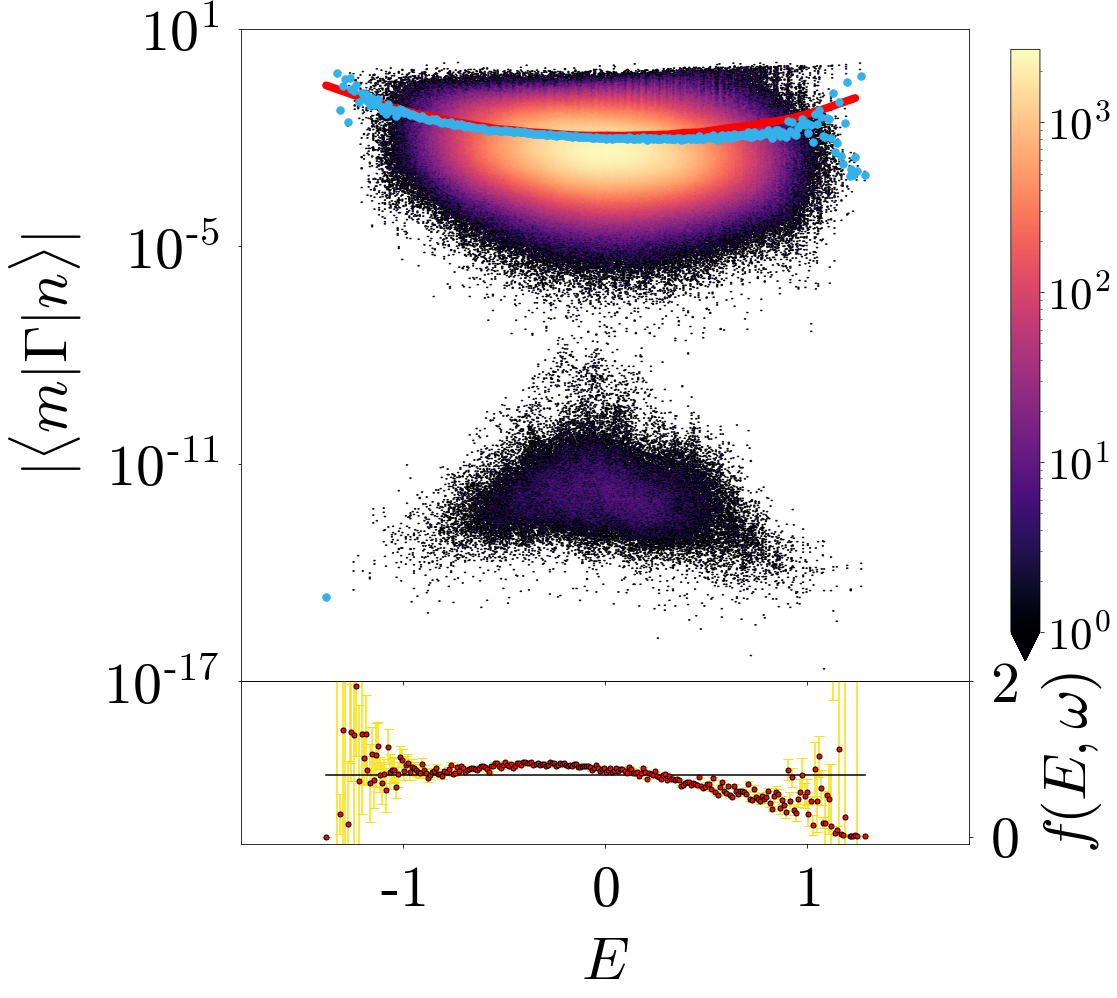}
\caption{Non-integrable}
\end{subfigure} &
\begin{subfigure}[r]{0.33\textwidth}
\includegraphics[width=\textwidth]{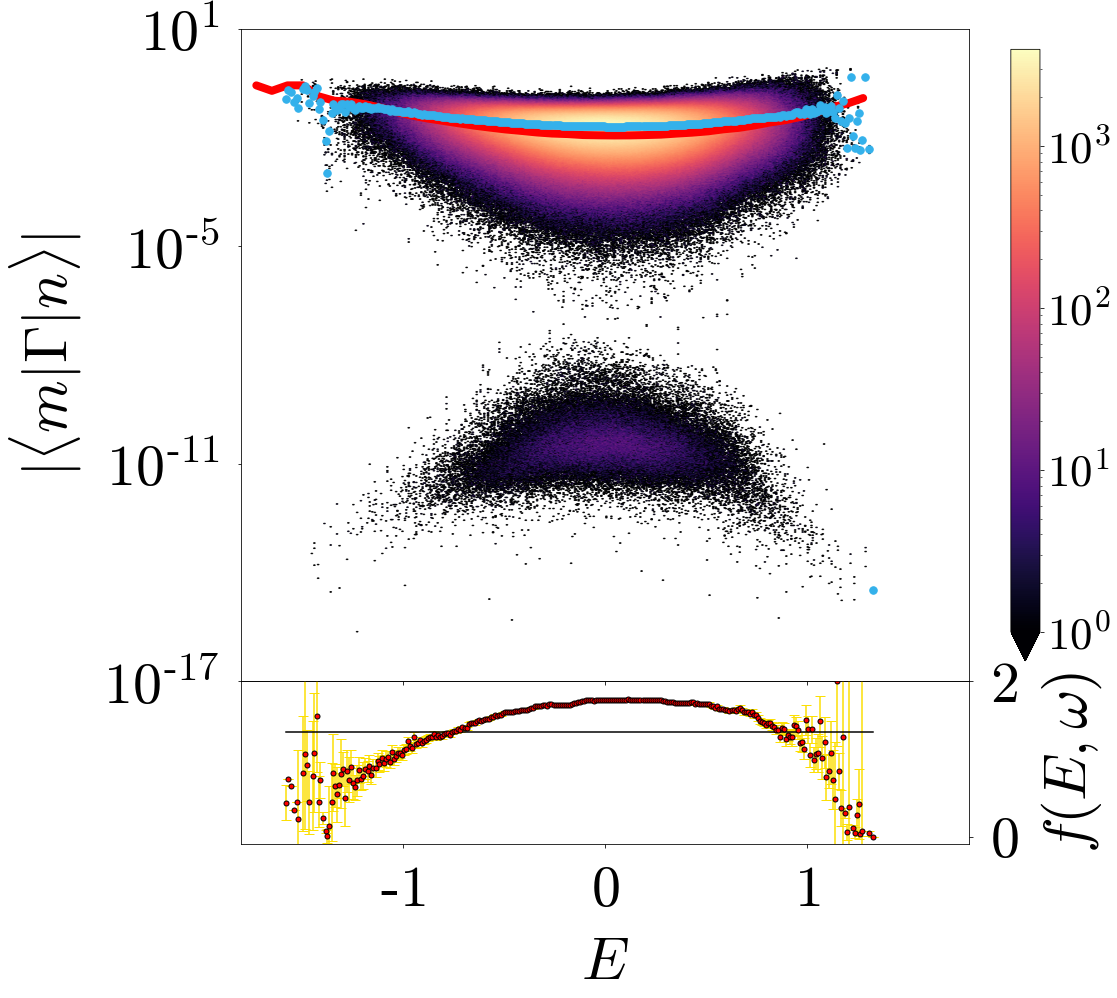}
\caption{Chaotic}
\end{subfigure}
\end{tabular}
\caption{(Colour online) Dependence of the absolute value of matrix elements of $\Gamma$ on the average energy of the states at fixed $\omega$. Note that in the integrable case, we do not perform an average or use a log scale, since the large majority of points are exactly zero. In the non-integrable and chaotic cases, the blue line corresponds to a running average over a small energy window and the red line is $e^{-S(E)/2}$. The bottom of the figure shows the running average divided by $e^{-S(E)/2}$, which gives the $E$ dependence of $|f(E,\omega)|$. Unlike for $\sigma^z$, there appears to be a non-trivial $E$ dependence. The error bars correspond to a 95\% confidence interval if the underlying distribution is normal.}
\label{fig:edep-gamma}
\end{figure}

By contrast, the matrix elements of the non-thermalizing operator $\Gamma$ clearly (by construction) do not follow an ETH-like distribution as a function of average energy or entropy in the integrable case. This is seen in figure \ref{fig:edep-gamma}. Indeed, they are very sensitive to the fine-grained structure of the states, as opposed to coarse-grained features like the average energy. However, immediately upon turning on the non-integrable deformation the matrix elements of the operator start to look ETH-like. From a microscopic point of view, this is not surprising, since there are no more details of the state for it to depend on: the additional conserved charges coming from integrability are at this point completely meaningless. We see here the effects of true ETH, which is able to overcome the fact that the operator was constructed in the integrable theory explicitly to evade OTH.
 There is one subtle distinction with the thermalizing operator $\sigma^z$. There is now a small remnant dependence on $E$ in addition to the entropic suppression $e^{-S(E)/2}$, i.e. the function $f(E,\omega)$ is not flat as a function of $E$. This dependence becomes more pronounced as the system becomes more chaotic. 

\subsection{Statistics}
\begin{figure}
\begin{tabular}[c]{cc}
\begin{subfigure}[c]{0.5\textwidth}
\includegraphics[width=\textwidth]{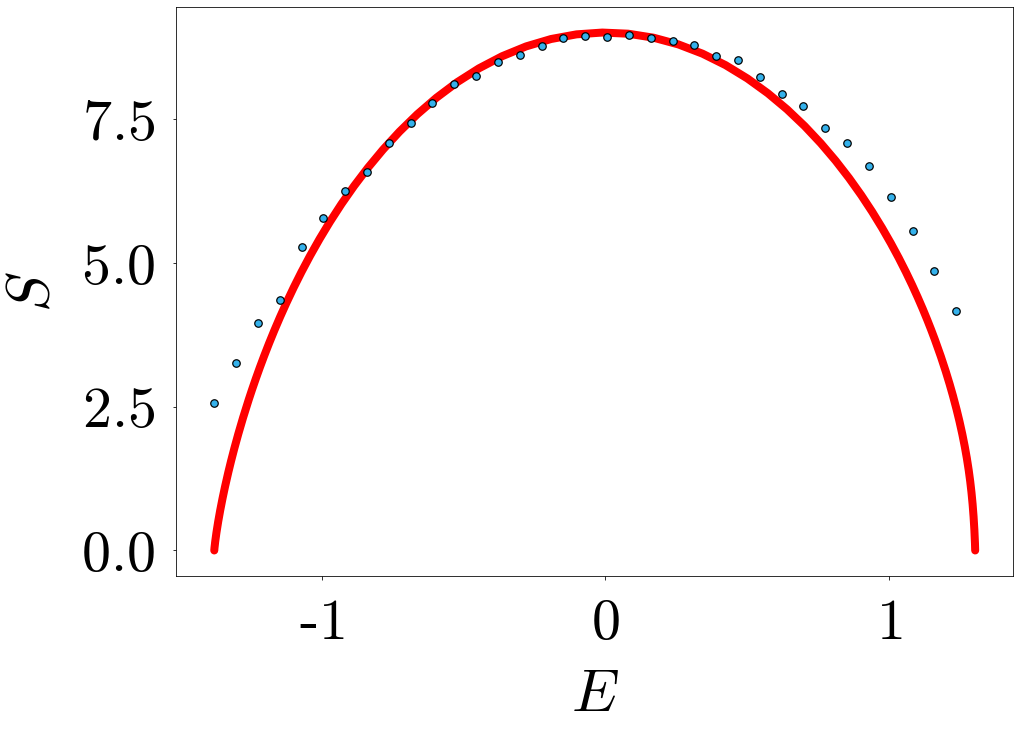}
\caption{Integrable}
\end{subfigure}&
\begin{subfigure}[c]{0.5\textwidth}
\includegraphics[width=\textwidth]{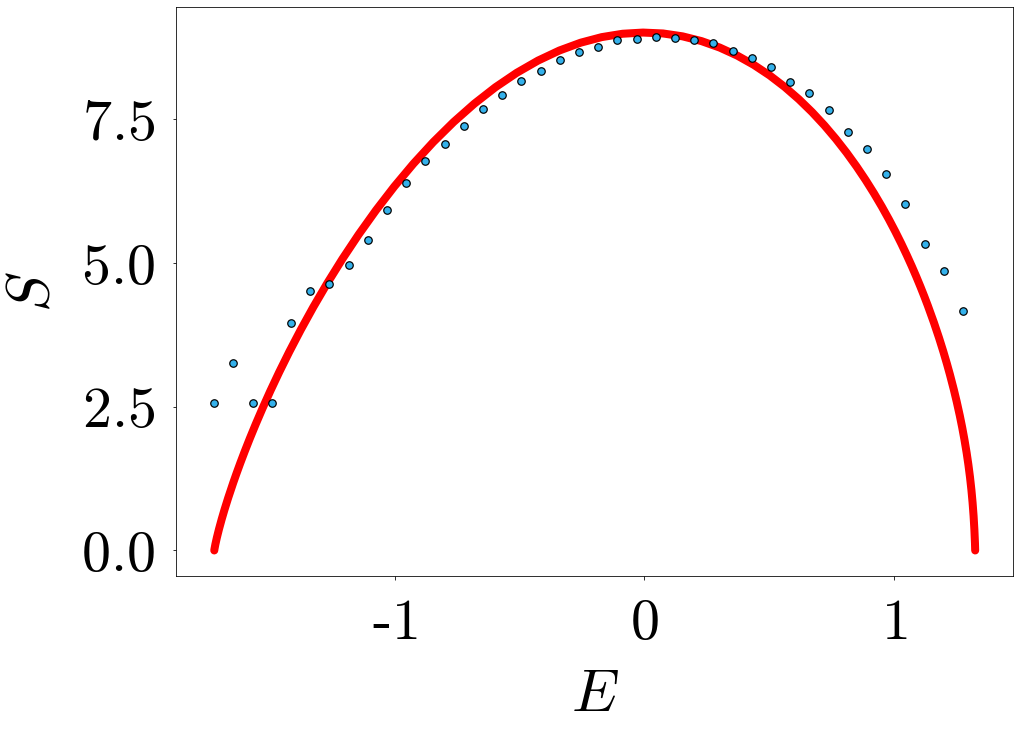}
\caption{Chaotic}
\end{subfigure}
\end{tabular}
\caption{(Colour online)Agreement between microcanonical and canonical entropy. The microcanonical entropy (blue points) is obtained from the logarithm of the number of states in a small but finite energy window (of size $\delta E=0.01$). The canonical entropy (solid red line) is obtained from the usual expression evaluated at a temperature where the average energy corresponds to the energy in question.}
\label{fig:entropies}
\end{figure}

We now probe a bit deeper into the meaning of the running average. As mentioned above, we extract $|f(E,\omega)|$ by averaging the magnitude of the matrix elements over a small window (where we assume this smooth function to be constant) and dividing the result by $e^{-S(E)/2}$, where $e^{S(E)}$ is the number of states in the window. The method to determine this entropy turns out to be irrelevant. Figure \ref{fig:entropies} shows the agreement between this microcanonical entropy and the usual canonical entropy calculated at a temperature set by the average energy. Note these two match best where the spectrum is densest (i.e. around $E=0$), and that in the integrable case it is essential to take a finite window. This is crucial. In order for the resemblance between OTH and ETH to become apparent, we have found that we must average over several energy levels in the integrable theory. A naive guess could have been that one only needed to sum over the degeneracy of a single energy level, and correspondingly in the non-integrable theory a window that just captured the splitting of these levels as the symmetry protecting the degeneracy is broken by the non-integrable deformation. This turned out to be insufficient and too narrow a window to see the resemblance between OTH and ETH. The resemblance is there for the larger window presented in figure \ref{fig:edep-sigma}.

\begin{figure}
\begin{tabular}[c]{ccc}
\begin{subfigure}{0.5\textwidth}
\includegraphics[width=\textwidth]{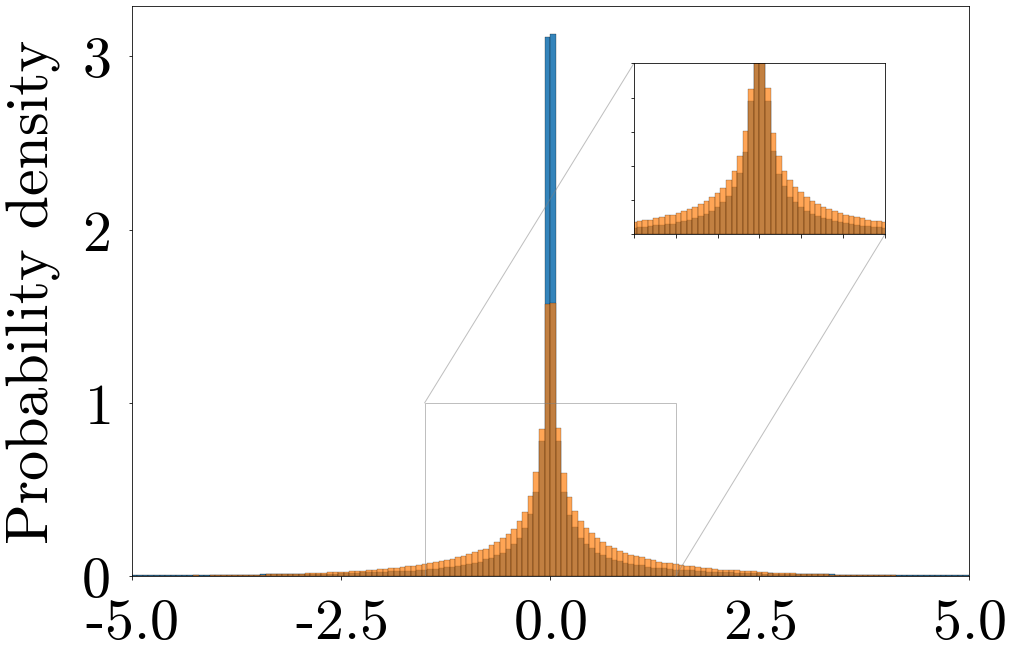}
\caption{$\text{Re}(R_{mn})$}
\end{subfigure} &
\begin{subfigure}{0.25\textwidth}
\includegraphics[width=\textwidth]{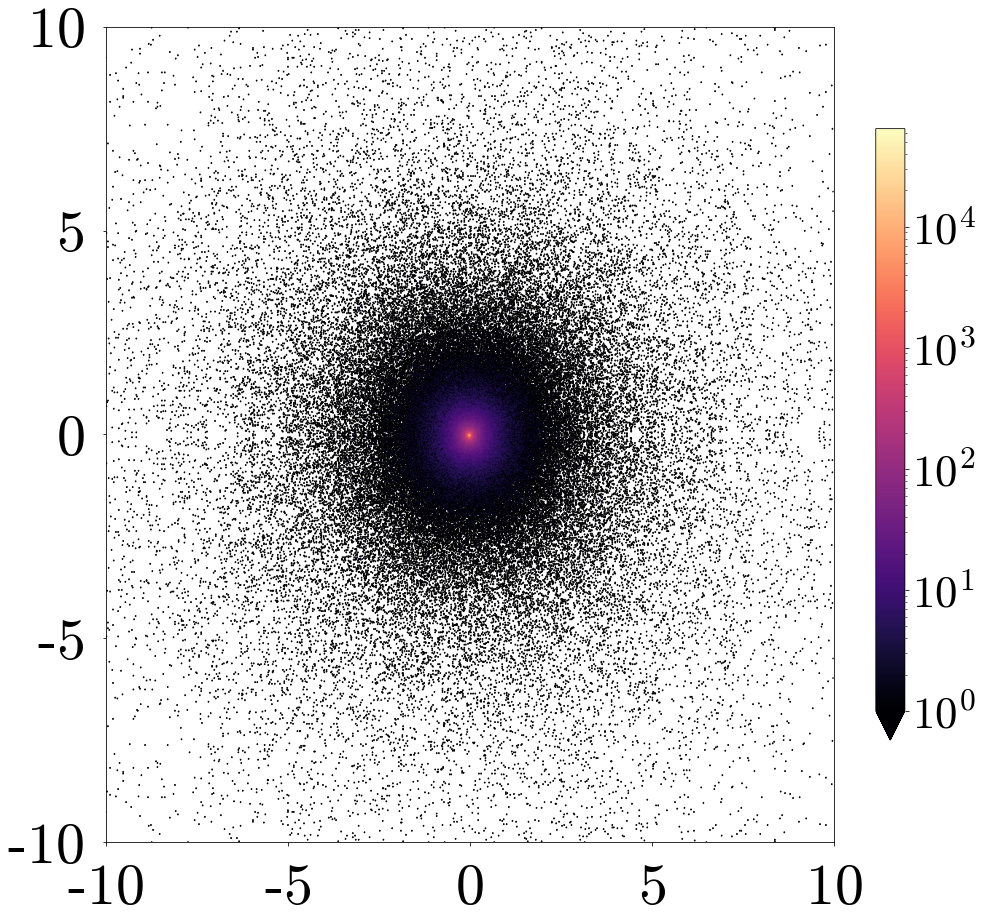}
\caption{$R_{mn}$ density in the complex plane, integrable ($h=0$)}
\end{subfigure} &
\begin{subfigure}{0.25\textwidth}
\includegraphics[width=\textwidth]{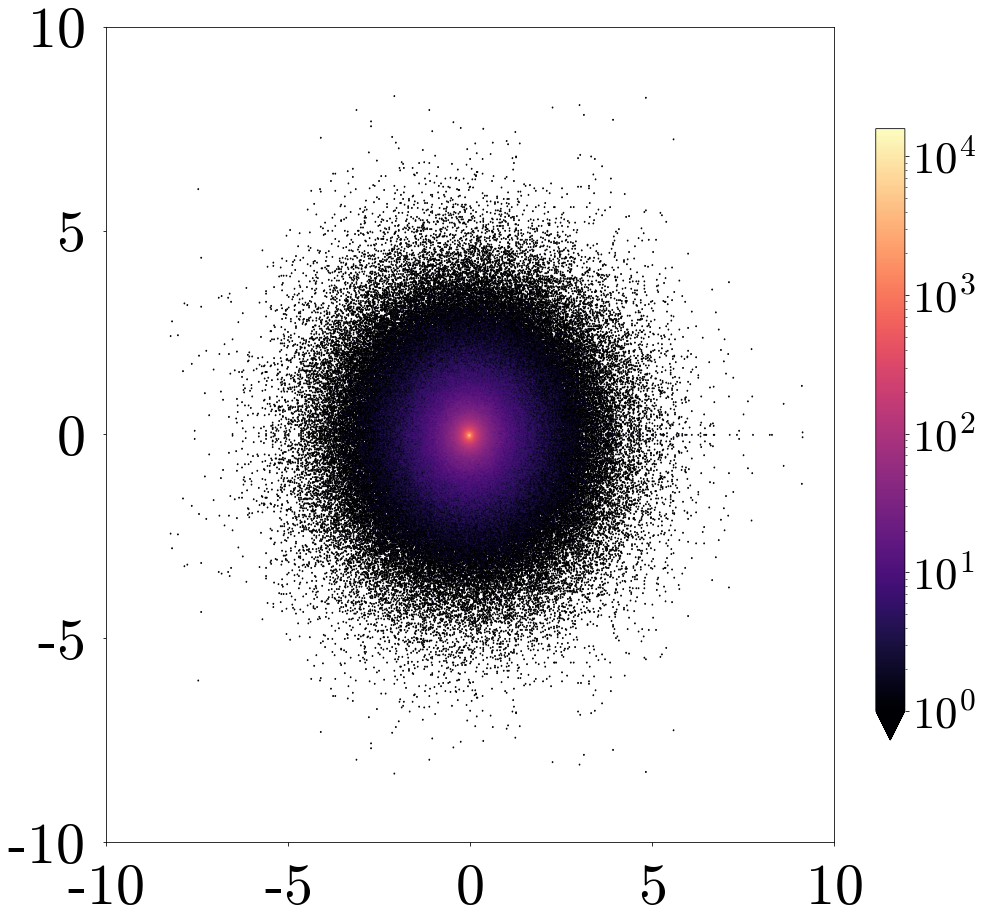}
\caption{$R_{mn}$ density in the complex plane, far from integrable ($h=0.5$)}
\end{subfigure} 
\end{tabular}
\caption{(Colour online) Statistical distribution of the matrix elements $e^{-S(E)/2} f(E,\omega) R_{mn}$ of $\sigma^z$ in a small energy window around $E=0$, $\omega=0$. $N=13$.}
\label{fig:stats-sigma}
\end{figure}

Insightful results follow from looking not at the average, but at the full statistical distribution. Following ETH, parametrize the matrix elements as
\begin{equation}
O_{mn}=e^{-S(E)/2} f(E,\omega) R_{mn}~.
\end{equation}
We now extract $R_{mn}$. In figure \ref{fig:stats-sigma}, we show a typical example of the distribution within a window of $\frac{O_{mn}}{\overline{|O_{mn}|}}$ for $\sigma_z$. Without loss of generality, we can take $\overline{|R_{mn}|}=1$ since $\overline{O_{mn}}=0$, so that $\frac{O_{mn}}{\overline{|O_{mn}|}}=R_{mn}$. 

\begin{figure}
\begin{tabular}[c]{ccc}
\begin{subfigure}{0.5\textwidth}
\includegraphics[width=\textwidth]{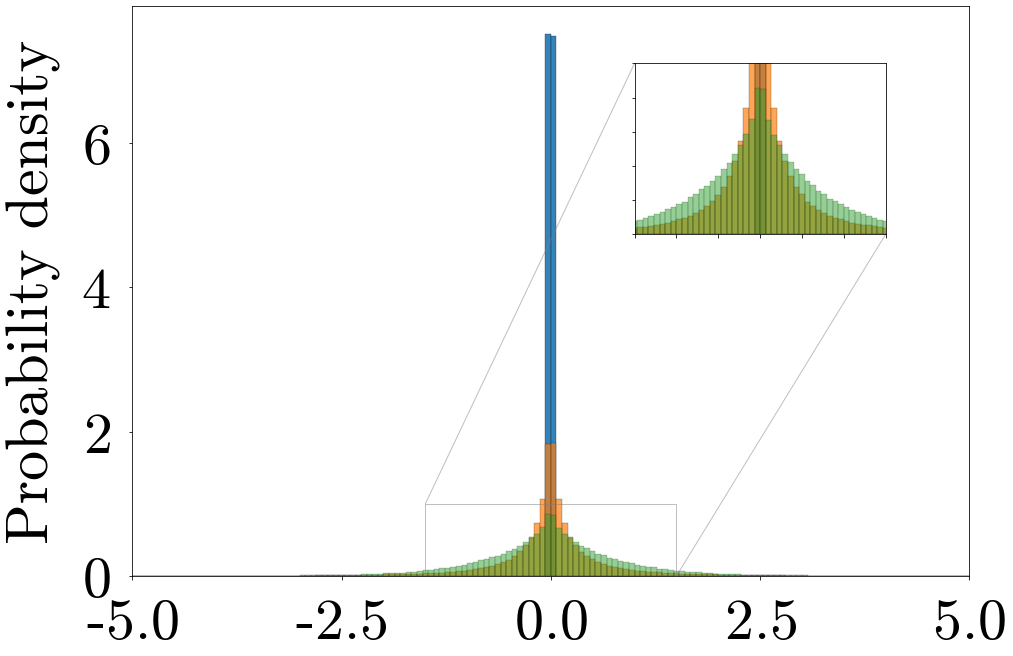}
\caption{$\text{Re}(R_{mn})$}
\end{subfigure} &
\begin{subfigure}{0.25\textwidth}
\includegraphics[width=\textwidth]{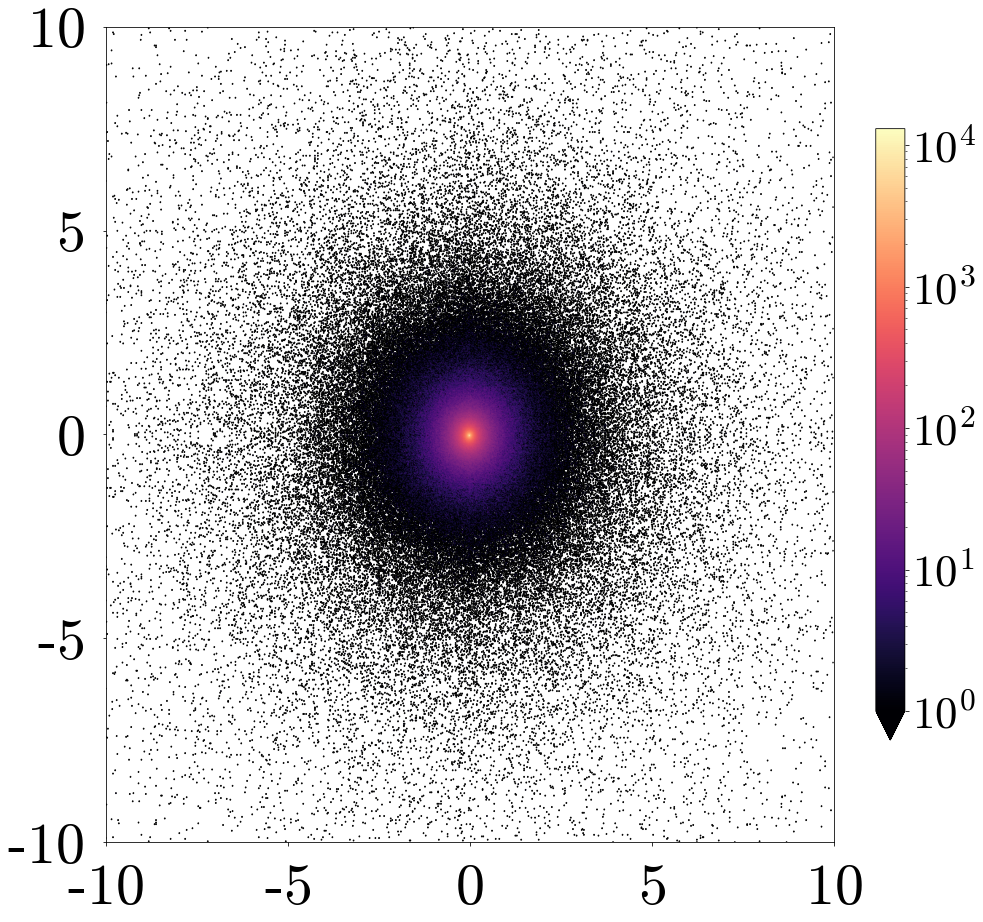}
\caption{$R_{mn}$ density in the complex plane, non-integrable ($h=0.1$)}
\end{subfigure} &
\begin{subfigure}{0.25\textwidth}
\includegraphics[width=\textwidth]{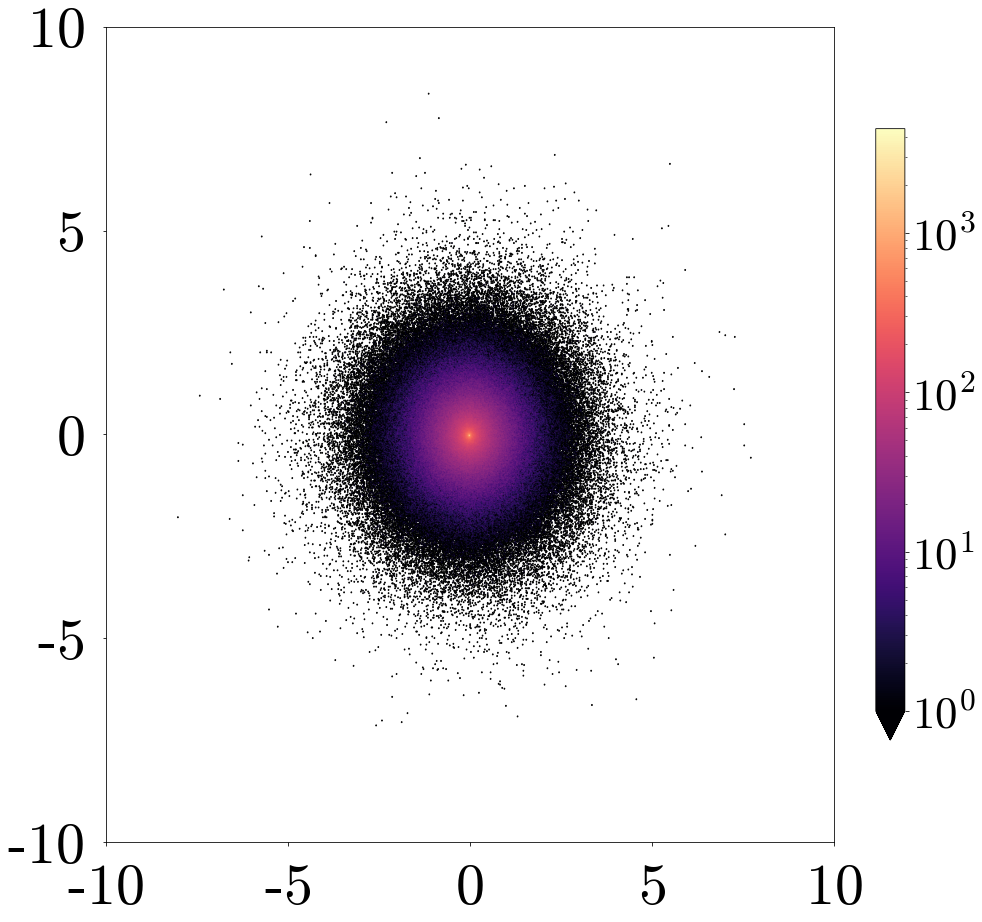}
\caption{$R_{mn}$ density in the complex plane, far from integrable ($h=0.5$)}
\end{subfigure} 
\end{tabular}
\caption{(Colour online) Statistical distribution of the matrix elements $e^{-S(E)/2} f(E,\omega) R_{mn}$ of $\Gamma$ in a small energy window around $E=0$, $\omega=0$. $N=13$.}
\label{fig:stats-gamma}
\end{figure}

\begin{figure}
\includegraphics[width=0.8\textwidth]{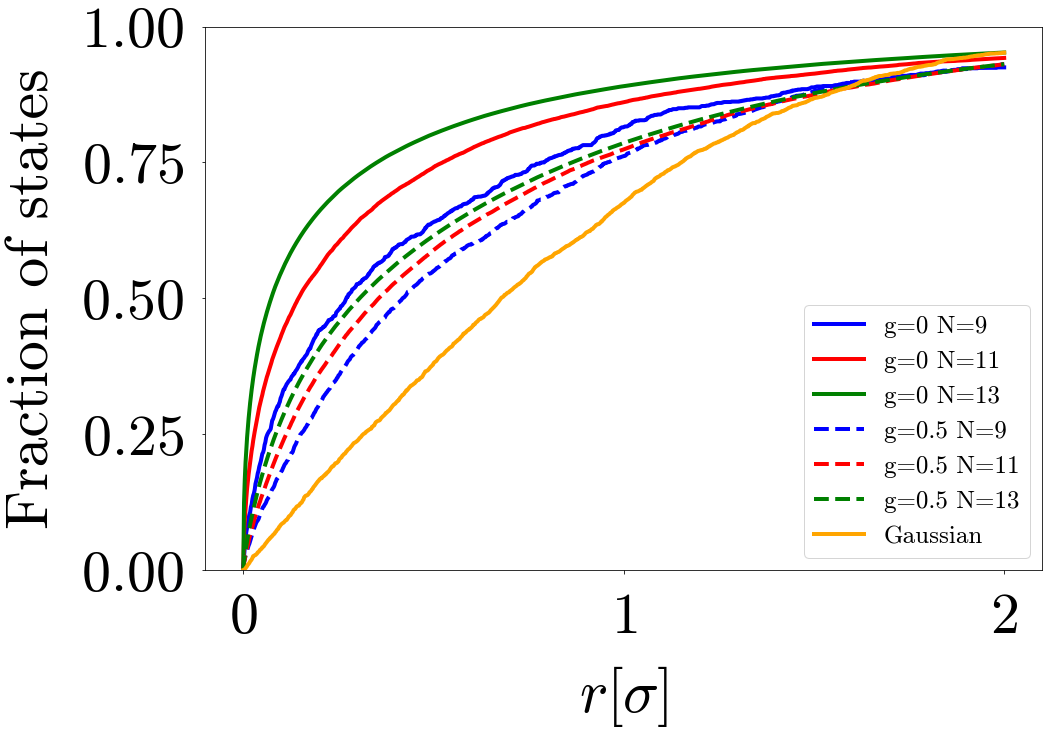}
\caption{The cumulative distribution function for $|R_{mn}|/\sigma$ (as defined in the text). A faster early growth corresponds to a more peaked distribution. A Gaussian is shown for reference. Notice that $\sigma^z$ in the integrable regime is not only more peaked than in the chaotic regime but becomes more so as $N$ increases.}
\label{fig:quantiles}
\end{figure}

We then see that the statistical distribution of values of $\sigma^z$ around the mean reveals a distinction between the integrable and non-integrable theories. In the non-integrable regime, where ETH should hold, the matrix elements in our energy window have a standard deviation of order 1. However, in the integrable regime the distribution is distinctly more peaked (although of comparable variance). To better understand this, we can plot the cumulative distribution of $|R_{mn}|$, that is to say
\begin{equation}
\chi(|R_{mn}|/\sigma)=\int_0^{|R_{mn}|/\sigma} P(|x|) dx~, 
\end{equation}
where $P(|x|)$ is the probability distribution of $R_{mn}/\sigma$. This is the probability that the absolute value of the matrix element is less than or equal to a particular value $|R_{mn}|$, with everything expressed in units of the standard deviation. In figure \ref{fig:quantiles}, we show how for $\sigma^z$ this function increases more sharply near zero in the integrable regime than in the chaotic regime, indicating that the probability distribution is more peaked. 

Considering the statistical distribution around the average for the non-thermalizing operator $\langle m|\Gamma|n\rangle$ the fluctuations around the average are again non-Gaussian distributed. In the integrable system, manifestly so. There is in essence no distribution. As the system becomes more chaotic, a distribution develops which is somewhat more peaked than the peaked non-Gaussian distribution for the thermalizing operator. This is seen in figure \ref{fig:stats-gamma}.

\subsection{Dependence on energy difference $\omega$}

\begin{figure}
\begin{tabular}[c]{cc}
\begin{subfigure}{0.5\textwidth}
\includegraphics[width=\textwidth]{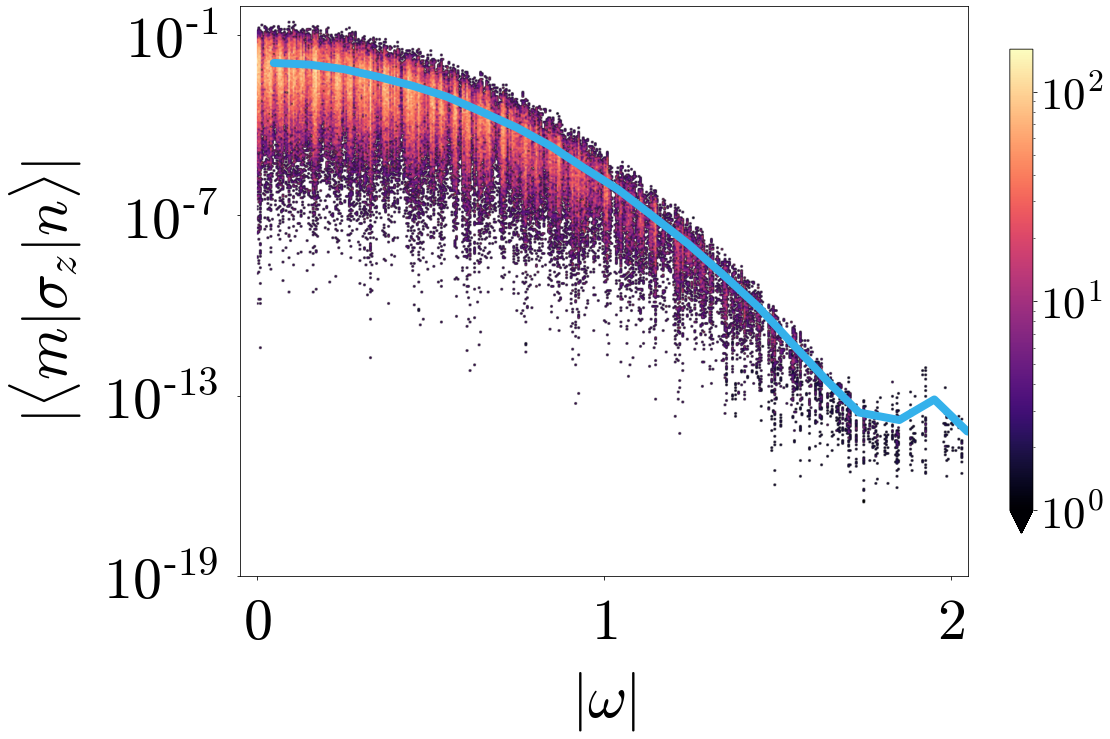}
\caption{Integrable}
\end{subfigure}&
\begin{subfigure}{0.5\textwidth}
\includegraphics[width=\textwidth]{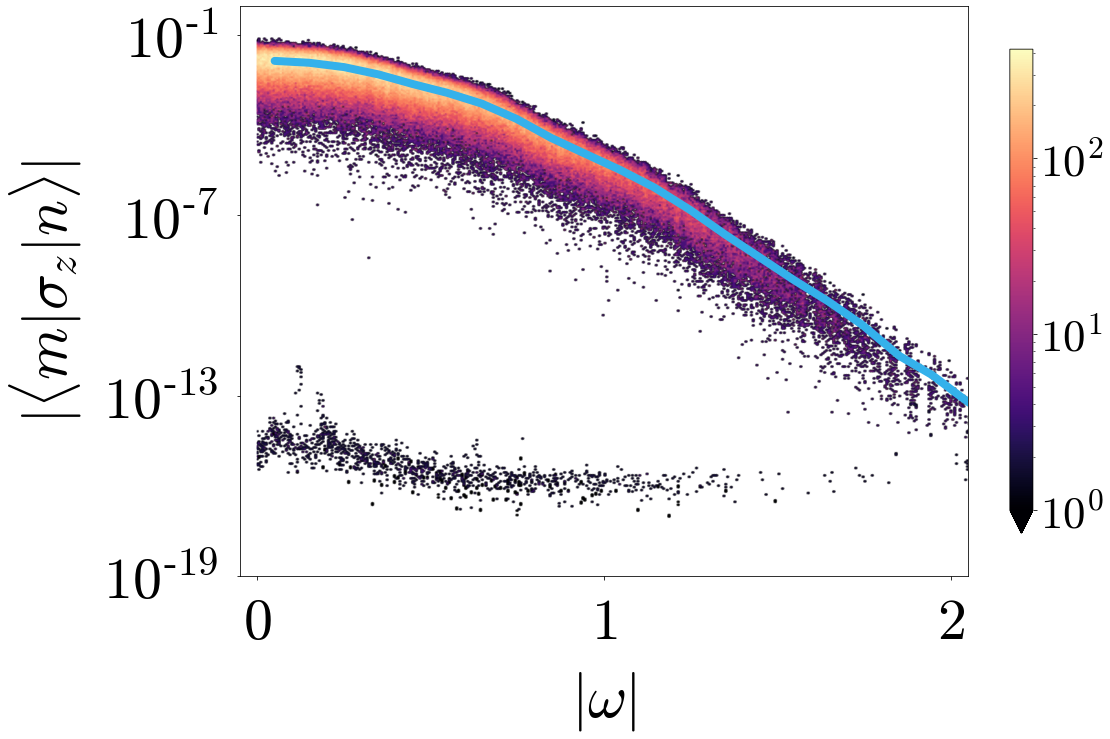}
\caption{Far from integrable}
\end{subfigure}
\end{tabular}
\caption{(colour online) Dependence of the absolute value of matrix elements of $\sigma^z$ on the energy difference between the states at fixed average energy $E$. The blue line corresponds to a running average over a small energy window, which is equivalent (up to an overall factor of $e^{-S(E)/2}$ to $|f(E,\omega)|$.}
\label{fig:wdep-sigma}
\end{figure}

The remaining ETH-like property to study is the dependence of the mean $f(E,\omega)$ on the energy difference. Fixing the average energy of the states and examining the dependence on the energy difference $\omega$, we again confirm the similarity between the matrix elements  $\langle m|\sigma^z|n\rangle$ of the thermalizing operator in the integrable and non-integrable theories. This is shown in figure  \ref{fig:wdep-sigma}. We show both the exact answer and a running average over a small energy window. The latter displays the expected smooth dependence of the matrix elements on energy---this time, the dependence on $\omega$ of $e^{-S(E)/2} f(E,\omega)$. A curious feature is that the dependence on $\omega$ is already noticeable at $\omega=0$. There is no random matrix theory-like plateau for $\omega<\omega_{*}$. Studying the energy difference dependence for $\Gamma$, on the other hand, does show this cut-off frequency below which the response is RMT-like, once the system has become non-integrable.\footnote{In the integrable regime, the dependence on $\omega$ is highly erratic: this is because the operator is very dependent on the details of the spectrum and cannot be simply understood in terms of the energies of the states.} We do not have an explanation for this distinction between the two operators. Nor does there appear to be a relation between $\omega_{*}$ and the relaxation time $\Omega^{-1}$ displayed in figure \ref{fig:life-damp}. We leave a better understanding of these scales to further study. We note, however, that the trivial $\omega$ dependence can be understood by the fact that $\Gamma$ loses all meaning when far away from integrability. Indeed it was built out of a few single-particle operators, but the physics of the model can no longer be understood in this language.

\begin{figure}
\begin{tabular}[c]{ccc}
\begin{subfigure}{0.33\textwidth}
\includegraphics[width=\textwidth]{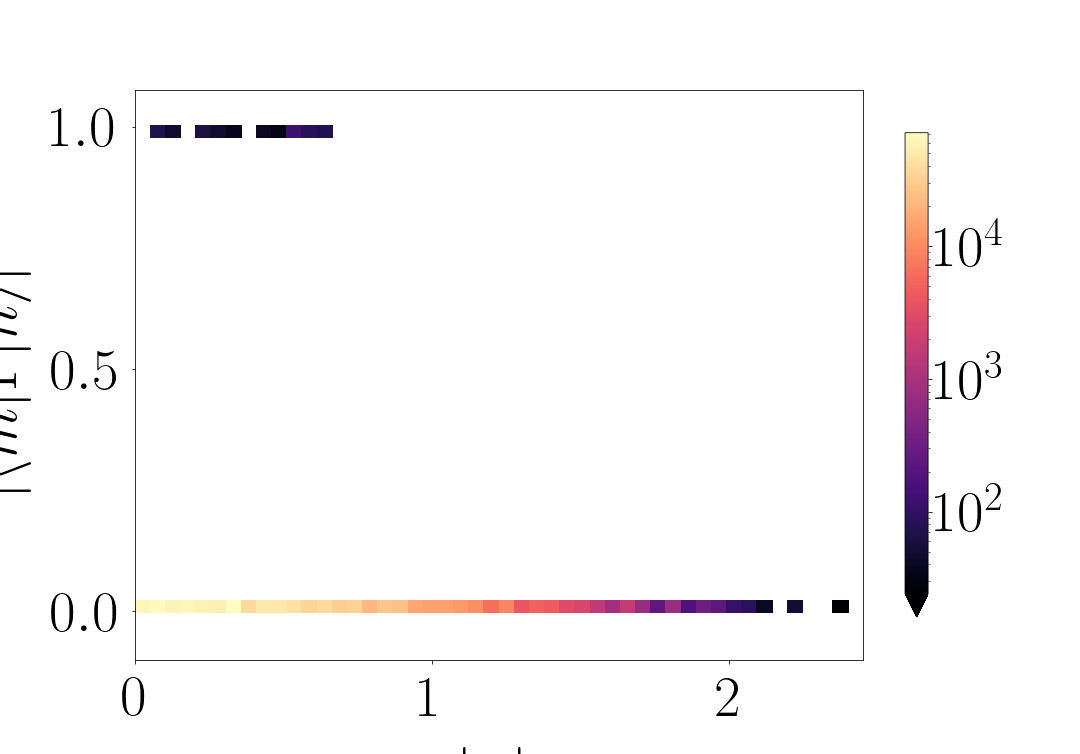}
\caption{Integrable}
\end{subfigure}&
\begin{subfigure}{0.33\textwidth}
\includegraphics[width=\textwidth]{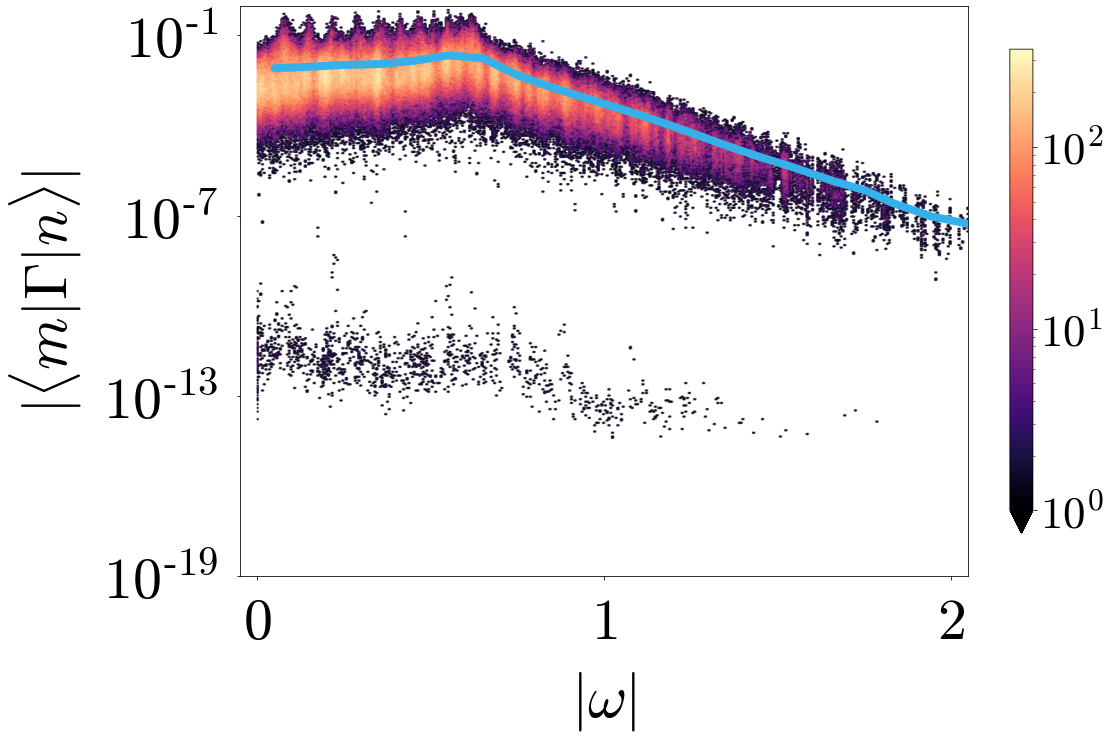}
\caption{Non-Integrable}
\end{subfigure}&
\begin{subfigure}{0.33\textwidth}
\includegraphics[width=\textwidth]{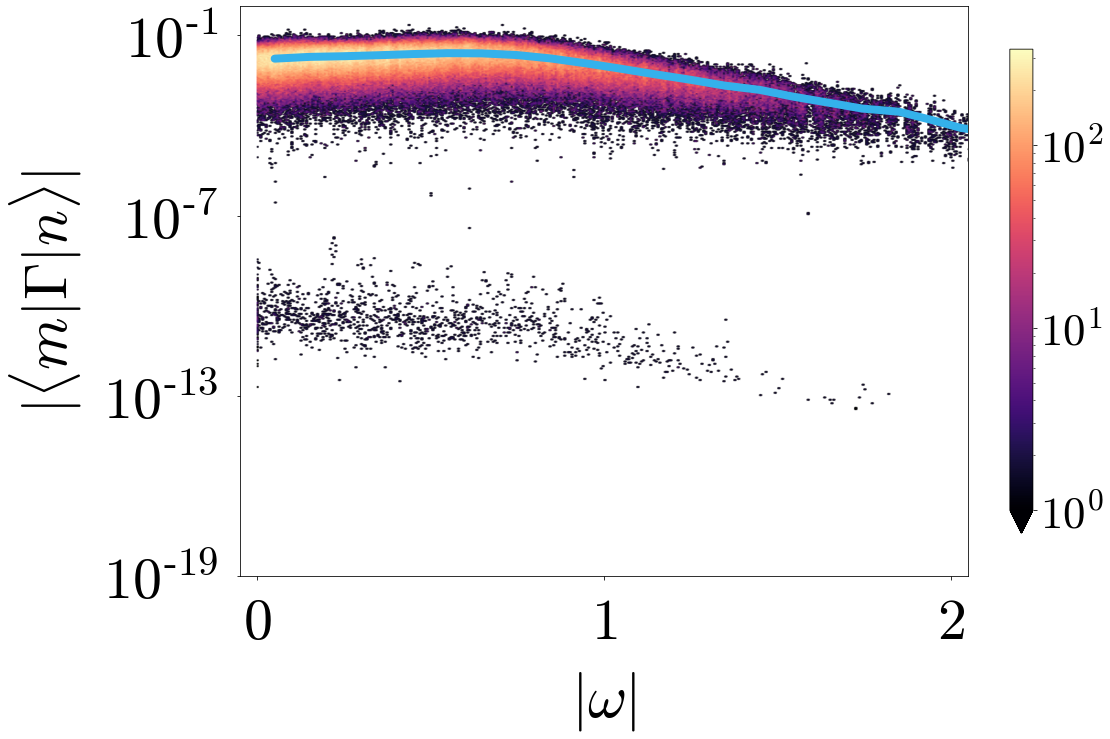}
\caption{Far from integrable}
\end{subfigure}
\end{tabular}
\caption{(Colour online) Dependence of the absolute value of matrix elements of $\Gamma$ on the energy difference between the states at fixed average energy $E$. The blue line in the last two figures corresponds to a running average over a small energy window, which is equivalent (up to an overall factor of $e^{-S(E)/2}$) to $|f(E,\omega)|$. $N=13$}
\label{fig:wdep-gamma}
\end{figure}

\section{Discussion and outlook}
In this note, we have demonstrated by explicit examples the differences and similarities between operator thermalization and eigenstate thermalization. We emphasize the point again: while the anstaz for matrix elements in equation (\ref{eq:eth-ansatz}) is such that operators that obey it will relax to their thermal expectation values, satisfying that ansatz for all (or most) operators in the theory is not necessary for there to be some operators that do relax. This is especially true when an average is taken, so that the details of the statistical distribution of matrix elements is smoothed over. Indeed, we have shown that $\sigma^z_i$ in a transverse field Ising chain, is consistent with this ansatz. This is despite the fact that the TFI is an integrable model. The corollary statement that an operator satisfying ETH implies quantum chaos is therefore also manifestly not true. 

We have also illustrated how the no-go condition is a feature of integrability: even a small move away from integrability caused our operator $\Gamma$ to relax, with the relaxation becoming faster as we moved farther away. This move away from integrability was also correlated with the matrix elements approaching a more ETH-like form.

A natural next route of inquiry is to study quenches and non-linear response. So far, we have focused on matrix elements and linear-response two-point functions. However, we can also ask how quenching with operators in different classes might produce states approximating different ensembles. An obvious question is whether there is a connection between operators satisfying the no-go condition and the resulting density matrices approaching a thermal ensemble vs a generalized Gibbs ensemble. This can be studied numerically using the examples we have presented here, but also analytically by closely examining the form that ETH should take in the presence of conserved charges and examining possible (in)compatibility with the no-go condition, equation (\ref{eq:no-go}).

Finally, one can wonder how far operators can go towards mimicking chaotic properties of the spectrum of theories, often the underlying physics behind ETH. This may be tested by studying the behaviour of out-of-time-order correlators and more generally operator growth. Once again, the example of thermalizing operators in free and integrable theories leads to the obvious question of whether they behave differently under such measures than non-thermalizing opertors. We hope to report on this soon.

\section*{Acknowledgements}
We are thankful to Marko Kuzmanovi\'{c}, Szilagyi D\'{a}niel, Tereza Vakhtel and Jan Zaanen for stimulating discussions. This research was supported in part by Koenraad Schalm’s VICI award of the Netherlands  Organization  for  Scientific  Research  (NWO),  by  the  Netherlands  Organization  for Scientific Research/Ministry of Science and Education (NWO/OCW), and by the Foundation for Research into Fundamental Matter (FOM). PSG acknowledges the support of the Natural Sciences and Engineering Research Council of Canada (NSERC) [PDF-517202-2018].
\bibliographystyle{jhep}
\bibliography{library}
\end{document}